\documentclass[trackchanges]{aastex701}

\defcitealias{Guo_Qiao_2026}{GQ26}

\usepackage{amsmath}
\usepackage[T1]{fontenc}
\usepackage[normalem]{ulem}
\usepackage{hyperref}   
\usepackage[toc,page]{appendix} 
\usepackage{amssymb}  

\makeatletter

\newcommand{\Rmnum}[1]{\expandafter\@slowromancap\romannumeral #1@}
\makeatother

\begin{document}

\title{Time-dependent Accretion Disks in Tidal Disruption Events: Long-term Light Curves}

\author[]{Chenlei Guo}
\affiliation{National Astronomical Observatories, Chinese Academy of Sciences, Beijing 100101, China; guocl@bao.ac.cn, qiaoel@nao.cas.cn}
\affiliation{School of Astronomy and Space Science, University of Chinese Academy of Sciences, Beijing 100049, China}
\email{guocl@bao.ac.cn}

\author[]{Erlin Qiao\thanks{Corresponding author}}
\affiliation{National Astronomical Observatories, Chinese Academy of Sciences, Beijing 100101, China; guocl@bao.ac.cn, qiaoel@nao.cas.cn}
\affiliation{School of Astronomy and Space Science, University of Chinese Academy of Sciences, Beijing 100049, China}
\email{qiaoel@nao.cas.cn}

\correspondingauthor{Erlin Qiao}
\email[show]{qiaoel@nao.cas.cn}



\begin{abstract}
The time-dependent accretion disk has been applied to explain the light curves observed in tidal disruption events (TDEs). Radiation pressure instability is expected to be an important factor that can shape the evolution of the accretion disk. In this paper, we upgrade the time-dependent disk model in Guo $\&$ Qiao by incorporating an index $\mu$ (with the stress tensor $\propto p^{\mu}p^{1-\mu}_{\rm{gas}}$, where $p=p_{\rm{gas}} + p_{\rm{rad}}$) for the modified viscosity, and a parameter $f_{\rm{w}}$ for the strength of the wind. Meanwhile, we adopt a more realistic fallback rate to inject into the disk. After systematically testing the effects of the newly incorporated parameters, 
we find that $f_{\rm{w}}$ can affect the time when the instability occurs, while $\mu$ can influence the occurrence and variation  magnitude of the radiation pressure instability. When $\mu<0.4$, the radiation pressure instability is completely removed from the disk, and the light curves evolve stably without large-scale magnitude variation. When $\mu>0.4$, the light curves can show oscillations caused by the instability, or drop steeply when the instability occurs and become flat in the late-time evolution, which mainly occurs when the viscous viscosity parameter $\alpha$ is small or the impact parameter $\beta$ is small. Since the drop magnitude can be modulated by $\mu$, we apply these `decay-to-flatten' light curves to some optical/UV observations in TDEs with different magnitudes of decline. Finally, we discuss the potential application of our model to some special TDEs that show oscillations in the light curves.

\end{abstract}


\section{Introduction}
{ A tidal disruption event happens when a star wanders too close to a super-massive black hole (BH). The tidal force of the BH exceeds the star's self gravity, tearing the star into pieces (\cite{Hills1975}, \cite{Rees1988}). About half of the stellar debris is ejected, while the other half can return back to the BH. The fallback debris dissipates orbital energy through collisions and eventually circularizes into an accretion disk. 
For $M_{\rm{BH}}\ge 10^{6}M_{\odot}$, a disk is expected to form within a few fallback timescales (\cite{Bonnerot2017MNRAS}). Then, the disk acts as a powerful source of energy, and its time-dependent evolution can strongly modulate the emission pattern in TDEs. So far, the time-dependent accretion disk has been applied to explain different light curves observed in TDEs (\cite{Mummery_Balbus2020MNRAS}, \cite{Luwenbin2022}, \cite{Alush_2025ApJ_1},  \cite{Piro_2025}, \cite{Guo_Qiao_2026}). 

Radiation pressure instability is expected to be one of the key components that can shape the evolutionary behavior of the accretion disk. This is a kind of thermal and viscous instability predicted to exist in a radiation pressure dominated thin disk (\cite{LE1974}), when a standard $\alpha-$prescription for viscosity is adopted. In TDEs, as the fallback rate decays from super-Eddington to be sub-Eddington on timescales of years, the radiation pressure instability may occur in the disk, 
which has been proposed to have potential applications in the observed light curves (\cite{Shen2014ApJ}, \cite{Yao_2022ApJ}, \cite{Luwenbin2022}, \cite{AT2020ocn_Pasham2024Natur}, \cite{Piro_2025}, \cite{Wu2026ApJ_instability}, \cite{Guo_Qiao_2026}, \cite{Daichi2026}).  
On the other hand, given that some of the light curves in TDEs evolve stably without showing large-magnitude variability, there are some time-dependent disk models in TDEs using different descriptions for viscosity or incorporating the magnetic pressure to remove the radiation pressure instability from the disk (\cite{Mummery_Balbus2020MNRAS}, \cite{Alush_2025ApJ_1}). These models are widely applied to explain the late-time optical/UV emissions in TDEs (\cite{Mummery2024MNRAS}, \cite{Mummery2025MNRAS.541..429M_Ref6},  \cite{Alush_2025_2}). This indicates that the exact description for the viscosity in the disk could vary from source to source. 

In \cite{Guo_Qiao_2026} (hereafter GQ26), we have calculated the evolution of a time-dependent disk in the standard TDE environment, i.e., injecting matter at the circularization radius of the stellar debris in the form of $\dot{M}_{\rm{inject}}\propto t^{-5/3}$. The basic model parameters include: BH mass $M_{\rm{BH}}$, viscosity parameter $\alpha$, and mass-injecting radius $R_{\rm{out}}$. The disrupted star is assumed to be a solar-type star with star mass $M_{\star}=M_{\odot}$ and radius $R_{\star}=R_{\odot}$. It was found that $\alpha$ and $R_{\rm{out}}$ can greatly affect the evolutionary behavior of the light curves. Particularly,  when $\alpha$ is small
or $R_{\rm{out}}$ is large (one can refer to Figure 3 and Figure 5 in GQ26 for details),
the oscillations predicted by the radiation pressure instability is completely suppressed. Instead, the light curves drop steeply when the instability occurs and become flat subsequently, which we used to explain the observed optical/UV light curves in some TDEs.
Hereafter, we refer to light curves with this behavior as `decay-to-flatten' light curves. 

In this work, we upgrade the basic disk model in GQ26 firstly by incorporating the modified viscosity.
Specifically, we model the stress tensor to be proportional to $p^{\mu}p_{\rm{gas}}^{1-\mu}$, with $\mu\in [0,1]$ the power law index, $p$ the total pressure and $p_{\rm{gas}}$ the gas pressure in the disk. This has been studied and widely used in the accretion disks in X-ray binaries and active galactic nuclei (AGN) (\cite{Taam_Lin_1984ApJ}, \cite{Szuszkiewicz1990MNRAS}, \cite{Honma1991PASJ}, \cite{Watarai_Mineshige_2003}, \cite{Merloni_Nayakshin2006_modified_visc}, \cite{Janiuk2017A&A_Modified_visc}). 
In addition, we incorporate the effect of disk wind in our disk model, which is suggested to be prominent in a super-Eddington accreting system (\cite{Jiang2014ApJ.796.106J}, \cite{Jiang2019ApJ.880.67J}, \cite{Dai2018ApJ}, \cite{Qiao2025MNRAS.539.3473Q}). We introduce a parameter $f_{\rm{w}}$ to describe the fraction of mass lost in the wind. In terms of the fallback rate, we adopt a more realistic form that contains the initial rising stage to inject into the disk. With all these newly incorporated factors, the parameters in our model are: $M_{\rm{BH}}$, the star mass $M_{\star}$ (and star radius $R_{\star}$), the impact parameter $\beta$, $\alpha$, $\mu$, $f_{\rm{w}}$ and $R_{\rm{out}}$, where $M_{\rm{BH}}$, $M_{\star}$($R_{\star}$) and $\beta$ are related to the disruption process of the star and the fallback rate. 

We systematically test the influence of all the newly incorporated parameters on the light curves. We find that $f_{\rm{w}}$ mainly affects the total luminosity and the time when the radiation pressure instability occurs. And $\mu$ can greatly modulate the variability amplitude magnitude induced by radiation pressure instability in the light curves. Meanwhile, $\mu$ determines whether the radiation pressure instability can occur in the disk. When $\mu<0.4$, we find that the instability is removed from the disk (see Figure \ref{Subplot_Scurve_mu} in Appendix \ref{appsec:Steady disk equations with outflows})}, which is also suggested previously by \cite{Kato2008book}. The light curves evolve stably without large-magnitude variation in this case. When $\mu>0.4$, the radiation pressure instability affects the evolution of the accretion disk. The light curves can show oscillations, or drop steeply when the instability occurs and become flat in the late times (decay-to-flatten), depending on the value of $\alpha$ and $\beta$.

As examples, we apply our light curves with decay-to-flatten behavior to explain the long-term evolution of the optical/UV light curves in some TDEs together with a photosphere model. We find that the light curves with different magnitudes of decline can be explained by our model by adjusting the model parameters. In particular, the drop magnitude in the light curves is dominated by the value of $\mu$.

This paper is organized as follows. In Section \ref{Sec.Model}, we present our disk model and the fallback rate in TDEs. In Section \ref{Sec.Results}, we display our results and test the effect of the newly incorporated parameters on the light curve. In Section \ref{Sec.Application to observations}, we show the application of our model to optical/UV light curves. We present the discussion in Section \ref{Sec.Discussion} and conclusions are presented in Section \ref{Sec.Conclusion}.

\section{Model}
\label{Sec.Model}
\subsection{Basic equations of time-dependent accretion disk}
\label{Subsec.Disk model}
We upgrade the time-dependent disk model in GQ26 by incorporating the radiation-driven outflows and the modified viscosity. 
Specifically, the mass conservation equation is
\begin{equation}
    \frac{\partial \Sigma}{\partial t} = \frac{1}{2\pi r}\frac{\partial \dot{M}}{\partial r}-\dot{\Sigma}_{\rm{out}}, \label{mass conserv}
\end{equation}
where
\begin{equation}
    \dot{M} = -2\pi r\Sigma v_{\rm{r}}. \label{M_dot}
\end{equation}
The angular momentum conservation equation is
\footnote{We assume that the outflowing material carries away the specific angular momentum it had at the point of ejection, which can be appropriate for the radiation-driven outflows considered here (\cite{Knigge1999MNRAS}). }
\begin{equation}
    \frac{\partial }{\partial t}\left (\Sigma r^{2}\Omega \right ) = \frac{1}{2\pi r}\frac{\partial}{\partial r}\left (\dot{M}r^{2}\Omega \right ) + \frac{1}{2\pi r}\frac{\partial}{\partial r}(2\pi r^{2}T_{\rm{r\varphi}}) - \dot{\Sigma}_{\rm{out}}r^{2}\Omega. \label{angular momentum conserv}
\end{equation}
The energy conservation equation is
\begin{equation}
     \Sigma T \frac{d S}{dt}= Q^{+}_{\rm{vis}} - Q^{-}_{\rm{rad}}-Q^{-}_{\rm{out}}. \label{energy conserv1}
\end{equation} 
The equations (\ref{mass conserv}), (\ref{M_dot}), (\ref{angular momentum conserv}) and (\ref{energy conserv1}) are vertically integrated over the disk. $\Sigma$ is the surface density, $T$ is the temperature, $\Omega$ is the angular velocity, $v_{\rm{r}}$ is the radial velocity, and $S$ is the entropy. The term $\Sigma T \frac{dS}{dt}$ on the left hand side of equation (\ref{energy conserv1}) represents the advection of energy. $\dot{\Sigma}_{\rm{out}}$ and $Q^{-}_{\rm{out}}$ are the mass loss rate and the energy loss rate per unit surface area due to the wind, respectively. $Q^{+}_{\rm{vis}}$ is the viscous heating rate and $Q^{-}_{\rm{rad}}$ is the radiative cooling rate per unit surface area.

We adopt the Keplerian angular velocity in a Newtonian potential, so the angular velocity $\Omega=\Omega_{\rm{K}}=\sqrt{GM_{\rm BH}/ r^3}$, where $G$ is the gravitational constant and $M_{\rm BH}$ is the BH mass. In this work, we adopt the modified viscosity following \cite{Janiuk2017A&A_Modified_visc}. Then the vertically integrated $r\varphi$ component of the stress tensor can be written as
\begin{equation}
    T_{\rm{r\varphi}}=\int_{-H}^{H} -\alpha p^{\mu}p_{\rm{gas}}^{1-\mu} dz=-2C_{1}\alpha p_{\rm{e}}^{\mu}(p_{\rm{gas,e}})^{1-\mu} H, \label{Trphi}
\end{equation}
where $\alpha$ is the dimensionless viscosity parameter, $\mu \in[0,1]$ is a dimensionless index. $H$ is the half-thickness of the disk, $p_{\rm{e}}$ and $p_{\rm{gas},e}$ are the total pressure and gas pressure on the equatorial plane of the disk, respectively. $C_{1}$ (also $C_{2}$ and $C_{3}$ in the following equations) is a vertical integration coefficient, described in detail in Appendix A of GQ26.
The total pressure $p$ in the disk can be expressed as
\begin{equation}
    p = \frac{k}{\mu_{\rm{m}} m_{\rm{p}}}\rho T + \frac{1}{3}aT^{4}. \label{total pressure}
\end{equation}
where $\rho$ is the density of the accretion flow. We take the mean molecular weight to be $\mu_{\rm{m}}=0.6$ assuming the solar metallicity. 

The hydrostatic equilibrium in the vertical direction can be written as
\begin{equation}
    \frac{p_{\rm{e}}}{\rho_{\rm{e}}} = C_{3}\Omega_{\rm{K}}^{2}H^{2}, \label{hydrostatic equilibrium}
\end{equation}
where $\rho_{\rm{e}}$ is the density on the equatorial plane of the disk.

Combining equation (\ref{mass conserv}), (\ref{M_dot}) and (\ref{angular momentum conserv}), together with (\ref{Trphi}) and (\ref{hydrostatic equilibrium}),  we obtain the expression for the radial velocity $v_{\rm{r}}$ as 
\footnote{In deriving this equation, we adopt an alternative expression for $T_{\rm{r\varphi}}$ in terms of $\nu$ and $\Sigma$, i.e., $T_{\rm{r\varphi}}=-2C_{1}\alpha p_{\rm{e}}^{\mu}(p_{\rm{gas,e}})^{1-\mu}H=-2C_{1}\alpha p_{\rm{e}}H\beta_{\rm{g}}^{1-\mu}=-2C_{1}\alpha \rho_{\rm{e}} c_{\rm{s}}^{2}H\beta_{\rm{g}}^{1-\mu}
=-2C_{1}\sqrt{C_{3}}\beta_{\rm{g}}^{1-\mu}\frac{3}{2}\nu \Sigma \Omega_{\rm{K}} $. }
\begin{equation}
    v_{\rm{r}} = -\frac{6C_{1}\sqrt{C_{3}}}{\Sigma r^{1/2}}\frac{\partial}{\partial r}(\beta_{\rm{g}}^{1-\mu}\nu \Sigma r^{1/2}). \label{vr}
\end{equation}
Here, $\beta_{\rm{g}}\equiv  p_{\rm{gas}}/p$, and $\nu = (2/3)\alpha H c_{\rm{s}}$ is the kinematic viscosity, where $c_{\rm{s}}$ is the isothermal sound speed defined as $c_{\rm s}=\sqrt{p_{\rm{e}}/\rho_{\rm{e}}}$. 

The viscous heating rate per unit surface area (two sides) is
\begin{equation}
  Q^{+}_{\rm{vis}}=rT_{\rm{r\varphi}}\frac{d\Omega}{d \mathit{r}}= 2C_{\rm{1}}\frac{3}{2}\alpha p_{\rm{e}}^{\mu}(p_{\rm{gas,e}})^{1-\mu}H\mathit{\Omega}. \label{Q_vis}
\end{equation} 
The radiative cooling rate per unit surface area is,
\begin{equation}
     Q^{-}_{\rm{rad}} = 2 C_{2}\frac{4\sigma T^{4}}{3\kappa \Sigma}, \label{Qrad}
\end{equation}
where $\sigma$ is the Stefan-Boltzmann constant, and $\kappa$ is the electron scattering opacity with $\kappa=0.34 \,\rm{cm^{2}g^{-1}}$ adopted.

In order to calculate the entropy gradient, we use the first law of thermodynamics,
\begin{equation}
     TdS=de + pd(1/\rho),
\end{equation}
where $e$ is the internal energy per unit mass.  
The internal energy per unit volume is $\rho e = 3/2p_{\rm{gas}} + 3p_{\rm{rad}}$ with adiabatic index $\gamma=5/3$. The 
vertically integrated internal energy can be written as
\begin{equation}
    E = \frac{3}{2}\frac{k}{\mu_{\rm{m}} m_{\rm{p}}}\Sigma T + aT^{4}H. \label{internal energy E}
\end{equation}

The energy equation (\ref{energy conserv1}) can then be expressed as, 
\begin{equation}
    \frac{\partial E}{\partial t} + \frac{\partial(rE v_{\rm{r}})}{r\partial r} + \Pi\frac{\partial(r v_{\rm{r}})}{r\partial r}= Q^{+}_{\rm{vis}} - Q^{-}_{\rm{rad}}-Q^{-}_{\rm{out}}, \label{energy conserv2}
\end{equation} 
where $\Pi=2C_{1}p_{\rm{e}}H$. 

The mass loss rate per unit surface area is defined as
\begin{equation}
    \dot{\Sigma}_{\rm{out}}(r)=\frac{\dot{M}_{\rm{out}}(r)}{\pi r^{2}}, \label{Sigma_out}
\end{equation}
where $\dot{M}_{\rm{out}}$ is the outflow rate. 
The energy loss rate per unit surface area is
\begin{equation}
    Q_{\rm{out}}^{-}(r) =\dot{\Sigma}_{\rm{out}}(r)e_{\rm{e}}
    =\dot{\Sigma}_{\rm{out}}(r)\left ( \frac{3}{2}\frac{kT}{\mu_{\rm{m}} m_{\rm{p}}} + \frac{aT^{4}H}{\Sigma} \right ), \label{Q_out}
\end{equation}
where $e_{\rm{e}}$ is the internal energy per unit mass on the equatorial plane of the disk\footnote{Here $e_{\rm{e}}=(\rho e)_{\rm{e}}/\rho_{\rm{e}}=\frac{3}{2}\frac{k}{\mu m_{\rm{p}}} + \frac{aT_{\rm{e}}^{4}}{\rho_{\rm{e}}}=\frac{3}{2}\frac{k}{\mu m_{\rm{p}}} + \frac{aT_{\rm{e}}^{4}H}{\rho_{\rm{e}}H}=\frac{3}{2}\frac{k}{\mu m_{\rm{p}}} + \frac{aT_{\rm{e}}^{4}H}{\Sigma}$, where $\Sigma\approx \rho_{\rm{e}}H$ (see details in Appendix A of GQ26). We use the disk temperature $T$ as an approximation for $T_{\rm{e}}$ }. 
The derivation of the angular momentum loss rate and energy loss rate in outflows can be found in Appendix \ref{appsec:Time-dependent disk equations with outflows}.

In modeling the radiation-driven outflow, we expect the wind to be strong when the disk is dominated by radiation pressure, and to become weak when the disk evolves to be gas-pressure-dominated. To automatically incorporate this variation of outflow strength during the disk evolution, we introduce 
$\beta_{\rm{rad}}=1-\beta_{\rm{g}}=p_{\rm{rad}}/p$ in modeling $\dot{M}_{\rm{out}}$. $\beta_{\rm{rad}}$ is not a model parameter, and is calculated by the derived $p$ and $p_{\rm{rad}}$ at each time step for each radius. Specifically, $\dot{M}_{\rm{out}}(r)$ can be expressed as 
\begin{equation}
    \dot{M}_{\rm{out}}(r)=f_{\rm{w}}\beta_{\rm{rad}}\dot M(r), \label{M_dot_out}
\end{equation} 
where $f_{\rm w}$ is assumed to be a constant and less than unity. For radiation pressure dominated state, $\beta_{\rm{rad}}\sim 1$, the ratio of the mass loss rate to the accretion rate is determined by $f_{\rm{w}}$. For gas pressure dominated state, $\beta_{\rm{rad}}\sim 0$, the outflow is almost negligible.


In summary, combining equation (\ref{total pressure}), (\ref{hydrostatic equilibrium}),  (\ref{vr}), (\ref{Q_vis}), (\ref{Qrad}), (\ref{internal energy E}), (\ref{Sigma_out}) and (\ref{Q_out}), we can solve equation  (\ref {mass conserv}) and (\ref {energy conserv2}) by specifying BH mass $M_{\rm{BH}}$, viscosity parameter $\alpha$, index $\mu$, wind parameter $f_{\rm{w}}$, the mass injecting rate $\dot{M}_{\rm inject}$, and the radius $R_{\rm{out}}$ where the mass is injected. The method in solving these equations and the settings of the boundary conditions are the same as that in GQ26.

\subsection{Fallback rate in TDEs}
\label{Subsec.Fallback rate}
When a star with mass $M_{\star}$ and radius $R_{\star}$ moves close enough to a supermassive BH, the star is torn apart at the tidal disruption radius $R_{\rm{T}}=R_{\star}(M_{\rm{BH}} / M_{\star})^{1/3}$. The impact parameter of the star is defined as $\beta=R_{\rm{T}}/R_{\rm{p}}$, where $R_{\rm{p}}$ is the pericenter distance of the stellar orbit. 
In this work, in order to account for the disk evolution in the early times of TDEs when the fallback rate is rising, we adopt the numerical results of the fallback rate in \cite{Guillochon2013ApJ}. Their simulations
were run for a wide range of $\beta$, with fixed $M_{\rm{BH}}=10^{6}M_{\odot}$ and $M_{\rm{\star}}=M_{\odot}$. 
The fallback rate with different $M_{\rm{BH}}$ and $M_{\rm{\star}}$ can be obtained by exerting the scaling relationships on the numerical results that they have presented (\cite{Guillochon_2014}, \cite{Mockler2019ApJ}). 
The relation between $\dot{M}_{\rm{fb}}(t) $ and $M_{\rm{BH}}$, $M_{\star}$, $R_{\star}$ can be scaled as (\cite{Mockler2019ApJ})
\begin{equation}
    \dot{M}_{\rm{fb}}(t) \propto M_{\rm{6}}^{-1/2}m_{\star}^{2}r_{\star}^{-3/2}. \label{M_dot_fb_prop}
\end{equation}
\begin{equation}
    t(\dot{M}_{\rm{fb}}) \propto M_{\rm{6}}^{1/2}m_{\star}^{-1}r_{\star}^{3/2}, \label{t_fb_prop}
\end{equation}
where $M_{6}=M_{\rm{BH}}/(10^{6}M_{\odot})$, $m_{\star}=M_{\star}/M_{\odot}$ and $r_{\star}=R_{\star}/R_{\odot}$. Here $t(\dot{M}_{\rm{fb}})$ is the time of a given fallback rate. The scaling relations (\ref{M_dot_fb_prop}) and (\ref{t_fb_prop}) follow from the standard calculation of the fallback rate in TDEs (\cite{Rees1988}, \cite{Evans_Kochanek1989ApJ}, \cite{Rossi_2021SSRv}).



As the fallback rate could be greatly influenced by the equation of state of the stars, \cite{Guillochon2013ApJ} constructed stars as polytropes, and ran the simulations with polytropic index $\gamma$ set to be $5/3$ and $4/3$, respectively. Following \cite{Mockler2019ApJ}, we use the fallback rate of different polytropes for stars of different masses: we take stars with mass $\le0.3M_{\odot}$ and mass $\ge 22M_{\odot}$ to be represented by $5/3$ polytropes ($\gamma=5/3$), while stars with masses between 1 and $15M_{\odot}$ to be represented by $4/3$ polytropes ($\gamma=4/3$). For stars in the transition ranges, the fallback rate can be obtained by smoothly blending between the 4/3 and 5/3 polytropes, which is described in detail in \cite{Mockler2019ApJ}. We use \cite{Tout1996MNRAS} to get $R_{\star}$ for a given $M_{\star}$, which is valid for $M_{\star}\ge 0.1M_{\odot}$. 

The value of $\beta$ for the full disruption of a star depends on the stellar type. Defining $\beta_{\rm{d}}$ to be the critical $\beta$ for complete disruption, it has been found that for stars with $\gamma=4/3$ polytrope, $\beta_{\rm{d}}=1.85$, and for stars with $\gamma=5/3$ polytrope, $\beta_{\rm{d}}=0.9$ (\cite{Guillochon2013ApJ}). 

The fallback debris will be circularized through collisions or physics that are not yet clearly understood. According to angular momentum conservation, the circularization radius is calculated 
to be
\begin{equation}
    R_{\rm{c}}=2R_{\rm{p}}=1.4\times 10^{13}\beta^{-1} M_{6}^{1/3}m_{\star}^{-1/3} r_{\star}\rm{cm}. \label{Rc}
\end{equation}
In this paper, following GQ26, we calculate the structure and evolution of the accretion disk in a TDE environment by continuously injecting mass at the circularization radius $R_{\rm{c}}$ in the form of the fallback rate. In summary, this work differs from GQ26 mainly in two aspects: (1) The disk model is upgraded to incorporate the modified viscosity and the outflows; (2) A more realistic fallback rate with the early rising stage is adopted. 

\section{Results}
\label{Sec.Results}
In this section, we demonstrate the light curves calculated by our disk model in different parameter spaces. Specifically, we focus on the effect of our newly incorporated parameter, namely $\mu$, $f_{\rm{w}}$ and $\beta$, on the light curves. We calculate the disk luminosity with $    L=2\int_{R_{\rm{ISCO}}}^{R_{\rm{out}}}\sigma T_{\rm{eff}}^{4} 2\pi rdr$ (equation (20) in GQ26), where $T_{\rm{eff}}$ is the effective temperature of the accretion disk at some fixed radius, and can be calculated with the formula of $T_{\rm{eff}} =[{Q^{-}_{\rm{rad}}/2\sigma}]^{1/4}$. 
In GQ26, it has been found that $\alpha$ is a key parameter which controls the light curves to be oscillating type (large $\alpha$) or decay-to-flatten type (small $\alpha$). In the following, in order to study both of these types of light curves, we test the effect of $\mu$, $f_{\rm{w}}$ and $\beta$ by setting a large $\alpha$ (i.e. $\alpha=0.05$) and a small $\alpha$ (i.e. $\alpha=0.005$) respectively.

\subsection{Effect of $\mu$}
\label{Subsec. effect of mu}

\begin{figure*}[ht!]
\centering
\includegraphics[scale=0.6]{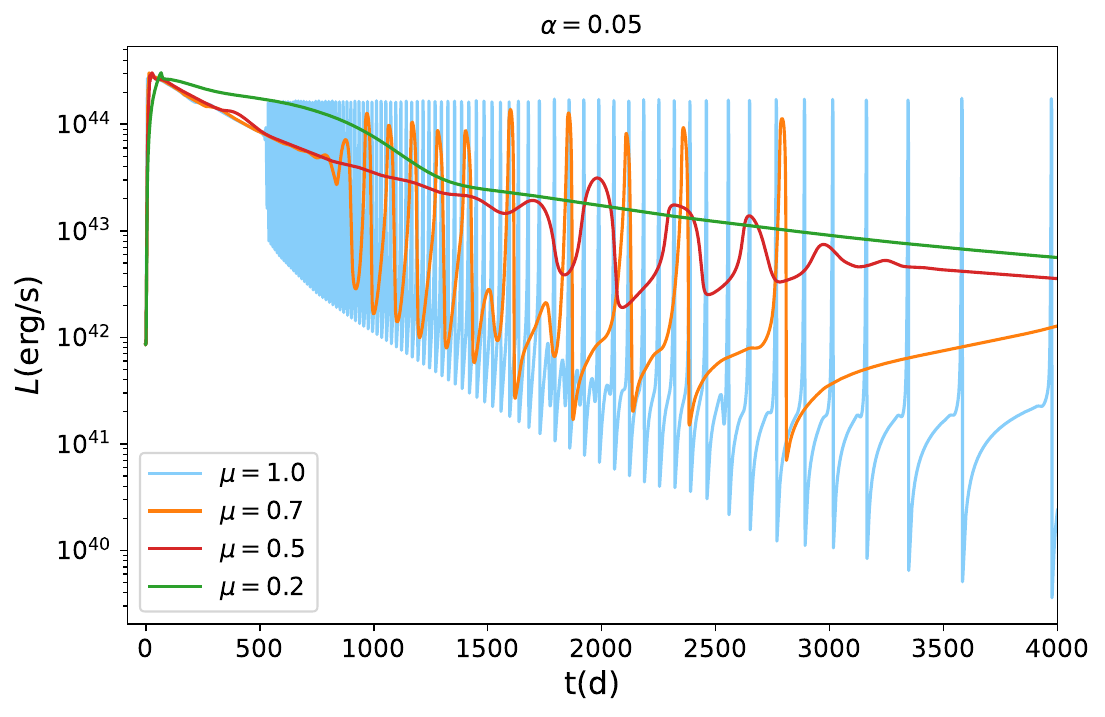}
\caption{ Light curves for different $\mu$, i.e., $\mu=1.0$, 0.7, 0.5, and 0.2. We set $M_{\rm{BH}}=10^{6}M_{\odot}$, $M_{\star}=M_{\odot}$, $\beta=\beta_{\rm{d}}=1.85$, $\alpha=0.05$, $f_{\rm{w}}=0$ and $R_{\rm{out}}=R_{\rm{c}}$ as equation (\ref{Rc}).
}
\label{Lc_mu_large_alpha}
\end{figure*}

\begin{figure*}[ht!]
\centering
\includegraphics[scale=0.4]{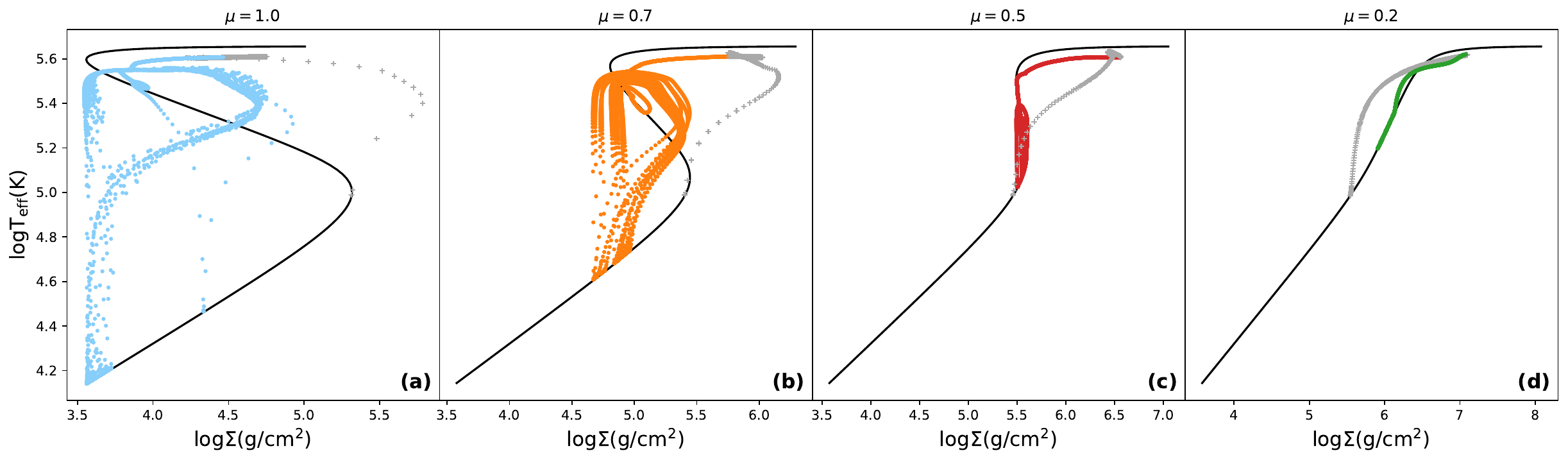}
\caption{ $T_{\rm{eff}}-\Sigma$ diagram for different $\mu$. Panels (a) to (d) are for $\mu=1.0$, 0.7, 0.5 and 0.2 respectively. The black S-shaped curve in the four panels are calculated with the thermal equilibrium equation (\ref{app:steady energy balance}) at $10R_{\rm{S}}$. The evolution of accretion disk in the $T_{\rm{eff}}-\Sigma$ diagram at $10R_{\rm{S}}$ is plotted in different colors for different $\mu$. The gray `+' indicate the initial rising stage of the light curves.
}
\label{Subplot_Scurve_mu_time}
\end{figure*}

In Figure \ref{Lc_mu_large_alpha}, we show the light curves calculated with $\alpha=0.05$ for different $\mu$, i.e., $\mu=1.0$, $0.7$, $0.5$ and $0.2$. For other parameters, we set $M_{\rm{BH}}=10^{6}M_{\odot}$, $\beta=\beta_{\rm{d}}=1.85$, $f_{\rm{w}}=0$ and $R_{\rm{out}}=R_{\rm{c}}$ as equation (\ref{Rc}). It can be seen that for $\mu=1.0$, $0.7$ and $0.5$, the light curves oscillate due to the radiation pressure instability. As $\mu$ decreases, the amplitudes of the oscillations become smaller, and the periods become longer. While for $\mu=0.2$, the light curve evolves stably throughout the 4000 days in our calculation without showing large-magnitude variability. In order to better demonstrate the effect of $\mu$, in Figure \ref{Subplot_Scurve_mu_time}, we show the corresponding evolution of the accretion disk in the $T_{\rm{eff}}-\Sigma$ diagram at $r=10R_{\rm{S}}$, with the same color standing for the same case in Figure \ref{Lc_mu_large_alpha}. Here $R_{\rm{S}}$ is the Schwarzschild radius defined as $R_{\rm{S}}\equiv 2GM_{\rm{BH}}/c^{2}$. The gray `+' indicates the initial rising stage of the light curves. For comparison, we plot the so-called S-shaped curves in black, which are calculated by solving the local energy balance equation (\ref{app:steady energy balance}). Panels (a) to (d) in Figure \ref{Subplot_Scurve_mu_time} are for $\mu=1.0$, $0.7$, $0.5$ and $0.2$, respectively. 

It can be seen that the unstable negative slope area on the S-shaped curve becomes smaller with smaller $\mu$. This is because a smaller $\mu$ corresponds to a smaller weight of $p_{\rm{rad}}$ in $T_{r\varphi}$ when the disk is dominated by $p_{\rm{rad}}$, as shown by equation (\ref{Trphi}), thus the radiation pressure instability is more strongly suppressed for smaller $\mu$. For $\mu=1.0$, $0.7$ and $0.5$, the radiation pressure instability still exists in the disk, but the limit-cycle becomes smaller for smaller $\mu$, leading to smaller amplitudes of the oscillations. Furthermore, for a radiation-pressure-dominated disk, $T_{r\varphi}$ decreases with decreasing $\mu$. Consequently, the viscous timescale becomes longer, resulting in longer periods of oscillations. For $\mu=0.2$, the negative slope area disappears, implying that the instability is completely removed from the disk. As shown in Figure \ref{Subplot_Scurve_mu}, this disappearance occurs for all $\mu<0.4$. In this case, the disk evolves stably along the thermal equilibrium curve, as can be seen in panel (d) of Figure \ref{Lc_mu_large_alpha}. The small change of slope in the light curve during the decaying phase is caused by the transition from a slim disk to a thin disk as the fallback rate decreases from super-Eddington to sub-Eddington.

For the $\mu=0.5$ case, it can be seen in Figure \ref{Lc_mu_large_alpha} that oscillation amplitudes can decrease over time, in contrast to the $\mu=1$ case where the peaks of the oscillations in the light curve remain unchanged. This can be understood as follows. For $\mu=1.0$, the disk drops to a low point on the lower stable branch of the S-shaped curve during each limit cycle (see panel (a) of Figure \ref{Subplot_Scurve_mu_time} for details). In this case, the disk must have accumulated enough materials to trigger the next outburst. Therefore, even though the mass supply rate decreases over time, $T_{\rm{eff}}$ can still jump to nearly the same peak value during each limit cycle, resulting in the same peak of luminosity in the light curve. For the $\mu=0.5$ case, the limit cycles are smaller. As the mass supply rate decreases, outbursts with smaller amplitudes can be triggered. As a result, the light curves for $\mu=0.5$ show a damping trend as the fallback rate decreases. 


\begin{figure*}[ht!]
\centering
\includegraphics[scale=0.6]{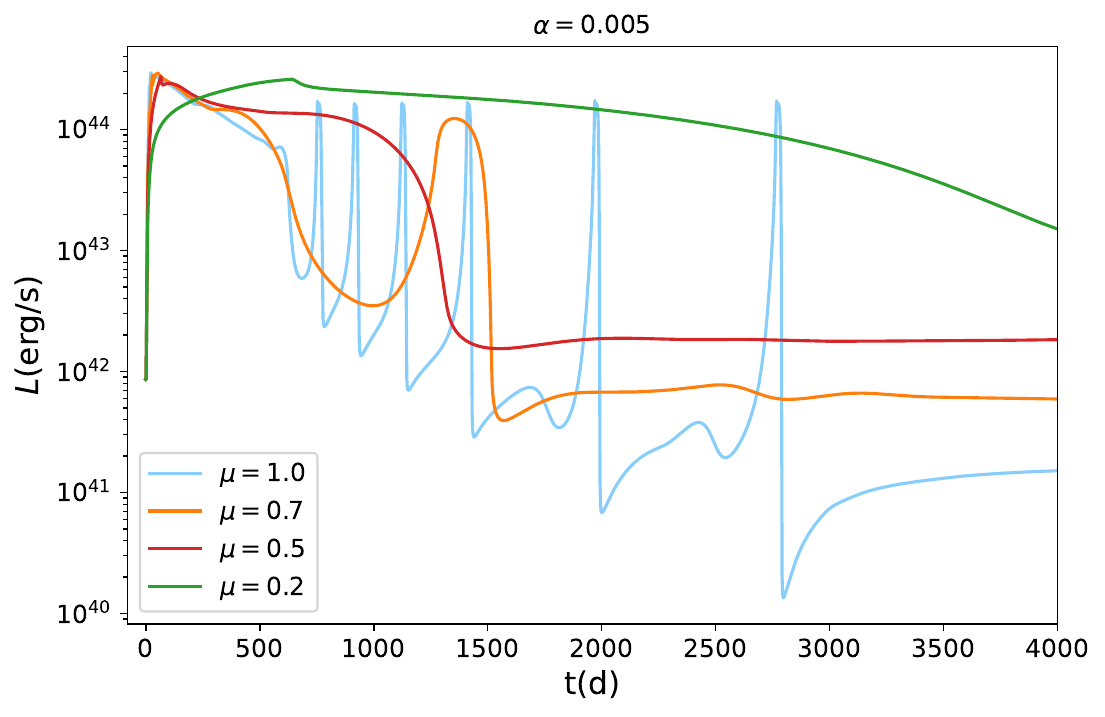}
\caption{Same as Figure \ref{Lc_mu_large_alpha} but with $\alpha=0.005$.}
\label{Lc_mu_small_alpha}
\end{figure*}

\begin{figure*}[ht!]
\centering
\includegraphics[scale=0.4]{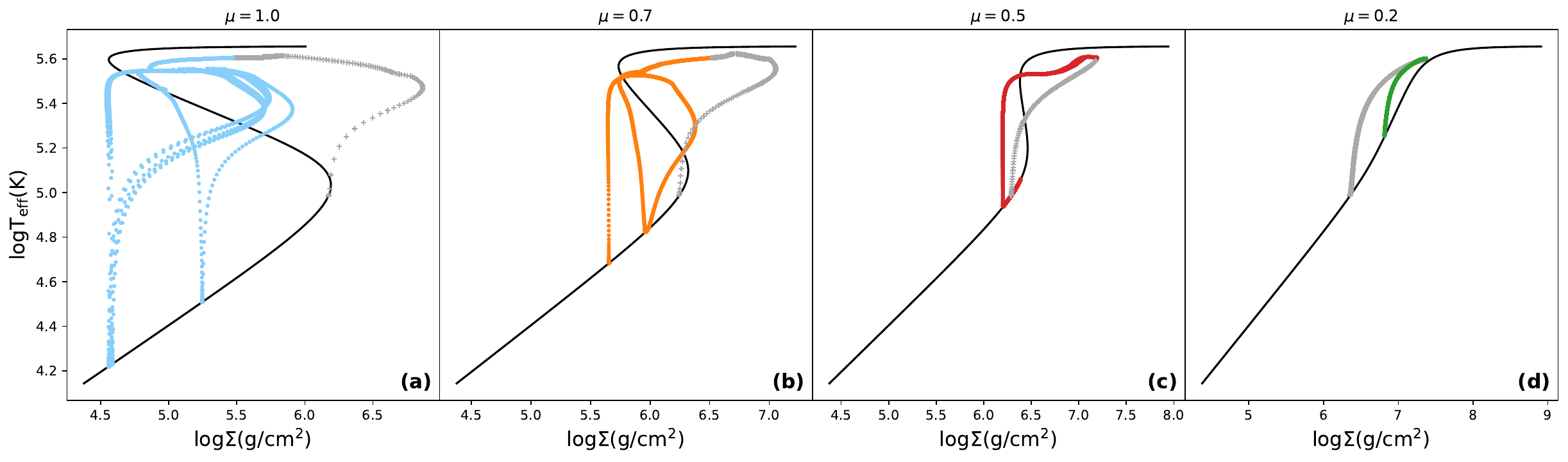}
\caption{ Same as Figure \ref{Subplot_Scurve_mu_time} but with $\alpha=0.005$.
}
\label{Subplot_Scurve_mu_time_alpha=0.005}
\end{figure*}

In Figure \ref{Lc_mu_small_alpha}, we show the light curves calculated with $\alpha=0.005$ for different $\mu$, i.e., $\mu=1.0$, $0.7$, $0.5$ and $0.2$. Other parameter settings are the same as in Figure \ref{Lc_mu_large_alpha}. Similarly, we also show the disk evolution on the $T_{\rm{eff}}-\Sigma$ diagram at $10R_{\rm{S}}$ for these four cases as those in Figure \ref{Subplot_Scurve_mu_time}. Compared with the $\alpha=0.05$ case shown above, it can be seen that for $\mu=1.0$ and $\mu=0.7$, the oscillating period becomes longer for $\alpha=0.005$. This is due to the longer viscous timescale caused by smaller $\alpha$. For $\mu=0.5$, the oscillations are completely suppressed. The light curve drops steeply when the radiation pressure instability occurs around $1200$ days, then becomes flat in late times. This decay-to-flatten behavior is similar to that found in GQ26 with small $\alpha$ or small $R_{\rm{out}}$ (see Figure 3 and Figure 5 in GQ26 for details). The difference is that the luminosity of the late-time plateau here is $\sim 10^{42} \rm{erg/s}$, which is higher than the $\sim 10^{41}\rm{erg/s}$ obtained with $\mu=1$ in GQ26. This enhancement occurs because a small $\mu$ partially suppresses the radiation pressure instability, shrinking the unstable region on the S-shaped curve, thus reducing the drop in $T_{\rm{eff}}$ when the instability occurs, as can be seen in panel (c) of Figure \ref{Subplot_Scurve_mu_time_alpha=0.005}. For the $\mu=0.2$ case, the decay in the light curve is shallower than that in Figure \ref{Lc_mu_large_alpha}, and the luminosity remains above $10^{43}\rm{erg/s}$ after the initial rise-to-peak until the end of the calculation at 4000 days. This slow evolution is also caused by the long viscous timescale with small $\alpha$ and small $\mu$.

\subsection{Effect of $f_{\rm{w}}$}
\label{Subsec. Effect of fw}

\begin{figure*}[ht!]
\centering
\includegraphics[scale=0.6]{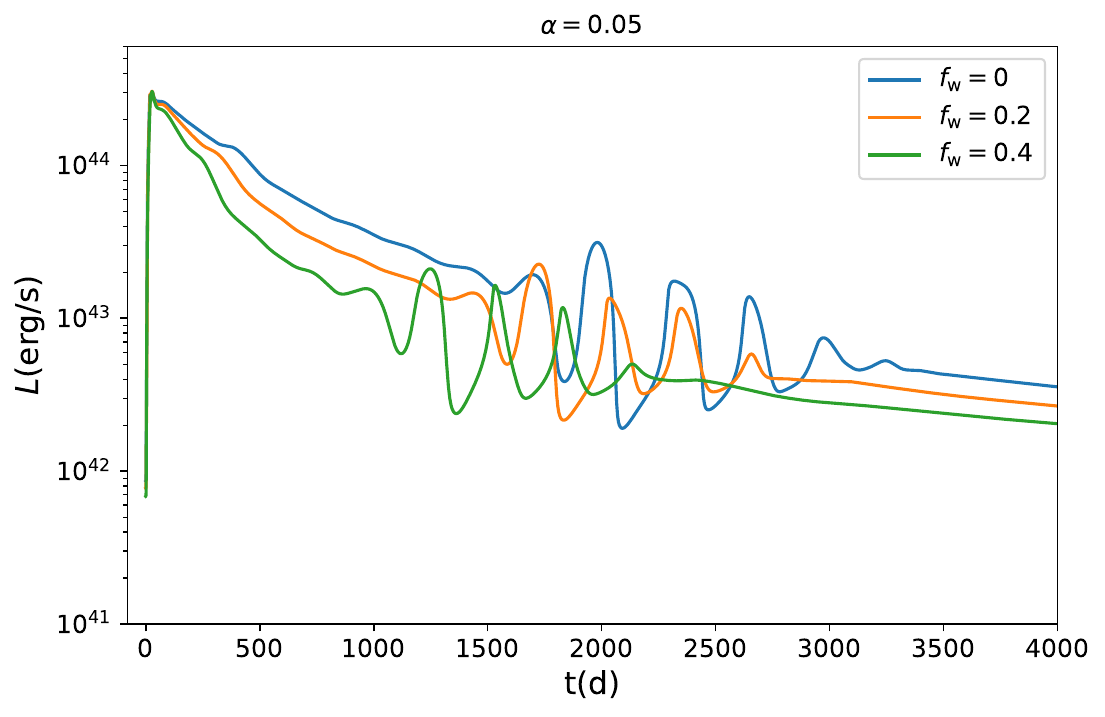}
\caption{  Light curves for different $f_{\rm{w}}$, i.e., $f_{\rm{w}}=0$, 0.2, and 0.4. We set $M_{\rm{BH}}=10^{6}M_{\odot}$, $M_{\star}=M_{\odot}$, $\beta=\beta_{\rm{d}}=1.85$, $\mu=0.5$, $\alpha=0.05$ and $R_{\rm{out}}=R_{\rm{c}}$ as equation (\ref{Rc}).
}
\label{Lc_fw_large_alpha}
\end{figure*}

\begin{figure*}[ht!]
\centering
\includegraphics[scale=0.6]{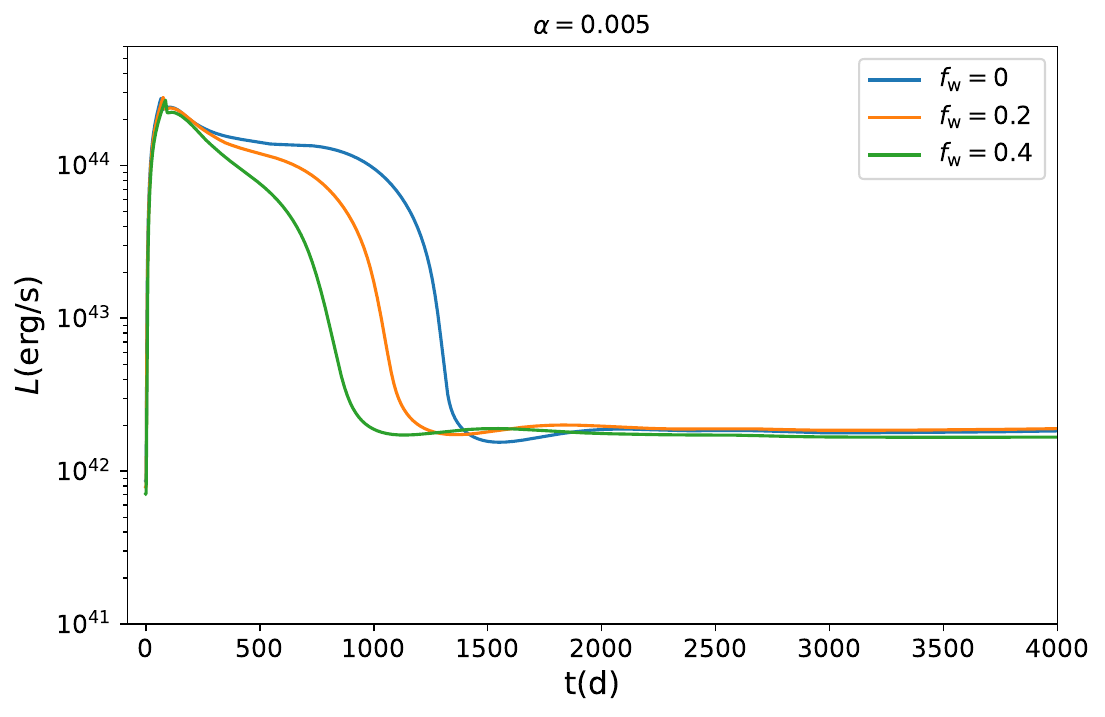}
\caption{ Same as Figure \ref{Lc_fw_large_alpha} but with $\alpha=0.005$.
}
\label{Lc_fw_small_alpha}
\end{figure*}

In Figure \ref{Lc_fw_large_alpha}, we plot the light curves calculated with $\alpha=0.05$ for different $f_{\rm{w}}$, i.e., $f_{\rm{w}}=0.0$, $0.2$ and $0.4$. We set $M_{\rm{BH}}=10^{6}M_{\odot}$, $\beta=\beta_{\rm{d}}=1.85$, $\mu=0.5$, and $R_{\rm{out}}=R_{\rm{c}}$ as equation (\ref{Rc}). Note that $f_{\rm{w}}$ determines the mass loss rate by outflow when the disk is dominated by radiation pressure, as shown in equation (\ref{M_dot_out}). It can be seen in Figure \ref{Lc_fw_large_alpha} that as $f_{\rm{w}}$ increases, the total luminosity decreases, and the oscillations caused by the radiation pressure instability occur earlier. This is because stronger outflows remove more materials from the disk, leading to smaller $\dot{M}$. With increasing $f_{\rm{w}}$, $\dot{M}$ will evolve to the critical value that triggers the instability earlier. Meanwhile, it can be seen that the total luminosity at the initial peak of the light curves is not significantly affected by $f_{\rm{w}}$. This is because the peak fallback rate is super-Eddington and a slim disk forms as the fallback rate rises to peak. For a slim disk, $L\propto \rm{ln}(\dot{M})$, so the luminosity of the disk varies weakly with the super-Eddington $\dot{M}$. As $\dot{M}$ decreases with the fallback rate, the influence of the outflows on the total luminosity gradually becomes apparent in the light curves. 

In Figure \ref{Lc_fw_small_alpha}, we plot the light curves calculated with $\alpha=0.005$ for different $f_{\rm{w}}$, i.e., $f_{\rm{w}}=0.0$, $0.2$ and $0.4$. Other parameter settings are the same as in Figure \ref{Lc_fw_large_alpha}. For small $\alpha$, the oscillations in the light curves are suppressed. The steep drop caused by the radiation pressure instability occurs earlier as $f_{\rm{w}}$ increases. The late-time luminosity is not affected by $f_{\rm{w}}$ in this case, since the wind driven by the radiation pressure that we model in this work automatically turns off when the disk transforms into a thin disk.

\subsection{Effect of $\beta$}
\label{Subsec. Effect of beta}
In this section, we test the effect of $\beta$ on the light curves with $\gamma=4/3$ and $\gamma=5/3$ types of stars separately. As specific examples, we take $M_{\star}=M_{\odot}$ for $\gamma=4/3$, and $M_{\star}=0.1M_{\odot}$ for $\gamma=5/3$.

\subsubsection{$M_{\star}=M_{\odot}$}
\label{Subsubsec 1Msun}

\begin{figure*}[ht!]
\centering
\includegraphics[scale=0.6]{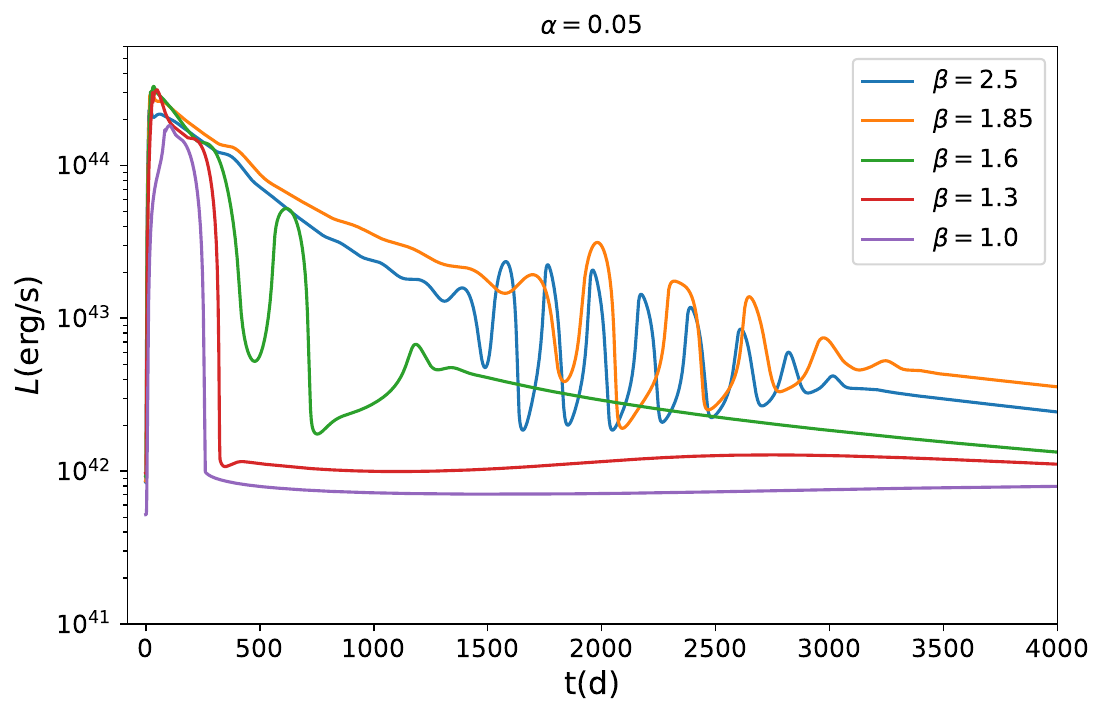}
\caption{ Light curves for different $\beta$, i.e., $\beta=2.5$, 1.85, 1.6, 1.3, and 1.0. We set $M_{\rm{BH}}=10^{6}M_{\odot}$, $M_{\star}=M_{\odot}$, $\mu=0.5$, $\alpha=0.05$, $f_{\rm{w}}=0$, and $R_{\rm{out}}=R_{\rm{c}}$ as equation (\ref{Rc}). 
}
\label{Lc_beta_1Msun_large_alpha}
\end{figure*}

In Figure \ref{Lc_beta_1Msun_large_alpha}, we plot the light curves calculated with $\alpha=0.05$ for different $\beta$, i.e., $\beta=2.5$, $1.85$, $1.6$, $1.3$, and $1.0$. We set $M_{\rm{BH}}=10^{6}M_{\odot}$, $\mu=0.5$, $f_{\rm{w}}=0$ and $R_{\rm{out}}=R_{\rm{c}}$ as equation (\ref{Rc}). For $\beta<\beta_{\rm{d}}$, i.e., the star is not fully disrupted, it can be seen that the oscillations are increasingly suppressed as $\beta$ decreases. This is because for shallower encounters, the fallback rate of a partial disruption event decreases more steeply with time  (\cite{Guillochon2013ApJ}). This rapid decay of the fallback rate leads to an earlier occurrence of the radiation pressure instability. Meanwhile, as the fallback rate quickly moves out of the unstable region of the S-shaped curve, the oscillations can only last for a very short time (e.g. $\beta=1.6$ case). For $\beta=1.3$ and $\beta=1.0$ cases, the fallback rate decays so rapidly that the oscillations are not triggered at all. The light curves also drop steeply when the instability occurs, and become flat in the late times. Compared to other decay-to-flatten light curves caused by small $\alpha$ in Figure \ref{Lc_mu_small_alpha} and Figure \ref{Lc_fw_small_alpha}, the light curves for small $\beta$ exhibit a steeper decay when the instability occurs, as can be seen in the red and purple light curves in Figure \ref{Lc_beta_1Msun_large_alpha}. 
This is because a small $\beta$ corresponds to a rapid decline in the fallback rate, which in turn causes a fast decay of $\dot{M}$ in the disk.

For $\beta\ge \beta_{\rm{d}}$, i.e., the star is fully disrupted, the light curves show a weaker dependence on $\beta$ compared to the partial disruption cases ($\beta<\beta_{\rm{d}}$). This is because the fallback rate remains largely similar across different values of $\beta$ for full disruptions(\cite{Guillochon2013ApJ}). The oscillations occur earlier for $\beta=2.5$, as its fallback rate declines slightly more steeply than that of the $\beta=1.85$ case. Moreover, the oscillation period of the $\beta=2.5$ case is shorter than that of the $\beta=1.85$ case. This is due to the decrease in $R_{\rm{c}}$ with increasing $\beta$, as shown by equation (\ref{Rc}).

\begin{figure*}[ht!]
\centering
\includegraphics[scale=0.6]{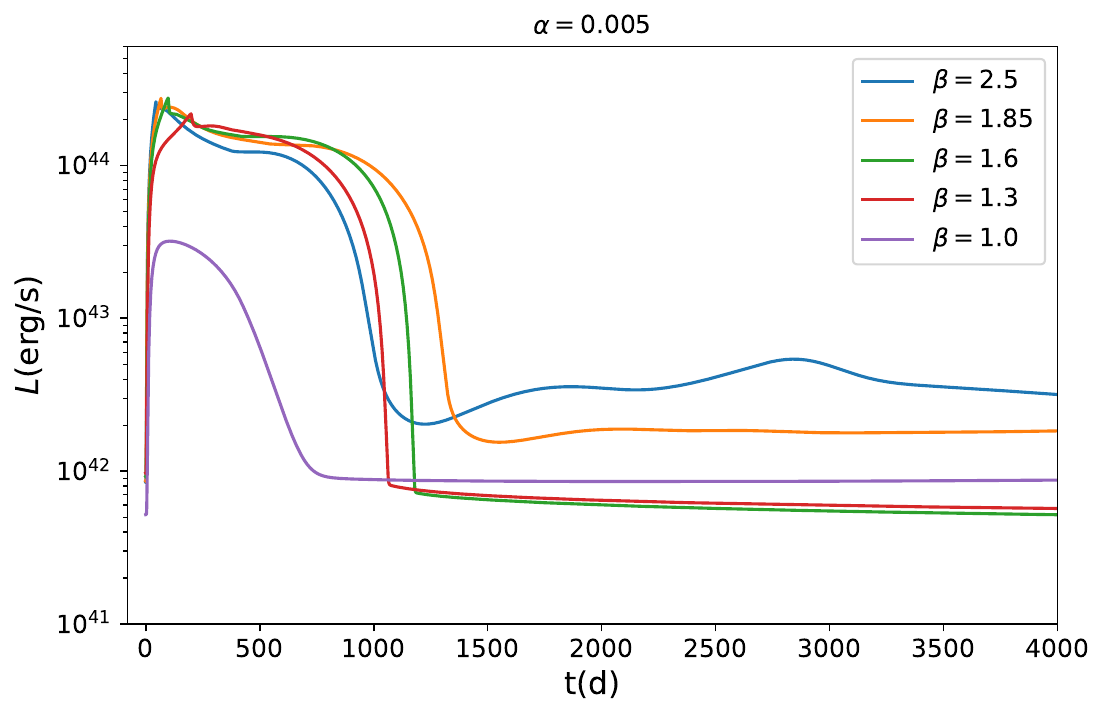}
\caption{ Same as Figure \ref{Lc_beta_1Msun_large_alpha} but with $\alpha=0.005$.
}
\label{Lc_beta_1Msun_small_alpha}
\end{figure*}

In Figure \ref{Lc_beta_1Msun_small_alpha}, we plot the light curves calculated with $\alpha=0.005$ for different $\beta$, i.e., $\beta=2.5$, $1.85$, $1.6$, $1.3$, and $1.0$. Other parameter settings are the same as in Figure \ref{Lc_beta_1Msun_large_alpha}. For $\beta<\beta_{\rm{d}}$, it can be seen that the steep drop caused by the instability occurs earlier with decreasing $\beta$, similar to the $\alpha=0.05$ case in Figure \ref{Lc_beta_1Msun_large_alpha}. The late-time luminosity varies with $\beta$, but in a nonlinear way. 
Specifically, the late-time luminosity decreases when $\beta$ decreases from 1.85 to 1.6, and remains nearly unchanged or slightly increase when $\beta$ decreases from 1.6 to 1.1. This nonlinear dependence arises because different $\beta$ corresponds to different properties of the fallback rate (including the peak of the fallback rate and the decay rate), while it does not modify the shape of the S-curve. A smaller $\beta$ can modulate the value of $T_{\rm{eff}}$ to a slightly lower value when the disk drops to the lower branch of the S-shaped curve, as the fallback rate has a lower peak value and decays more rapidly. However, the magnitude of this late-time $T_{\rm{eff}}$ is intrinsically determined by $\mu$ (see Figure \ref{Subplot_Scurve_mu}). Hence a smaller $\beta$ could cause a lower late-time luminosity, but only to a limited extent.

For $\beta >\beta_{\rm{d}}$, the $\beta=2.5$ case exhibits an earlier occurrence of the steep drop caused by the instability, and a small tendency of oscillation. This is a result of the slightly more steep decline in the fallback rate and the smaller $R_{\rm{c}}$ for $\beta=2.5$ as analyzed above. 

\subsubsection{$M_{\star}=0.1M_{\odot}$}
\label{Subsubsec 0.1Msun}

\begin{figure*}[ht!]
\centering
\includegraphics[scale=0.6]{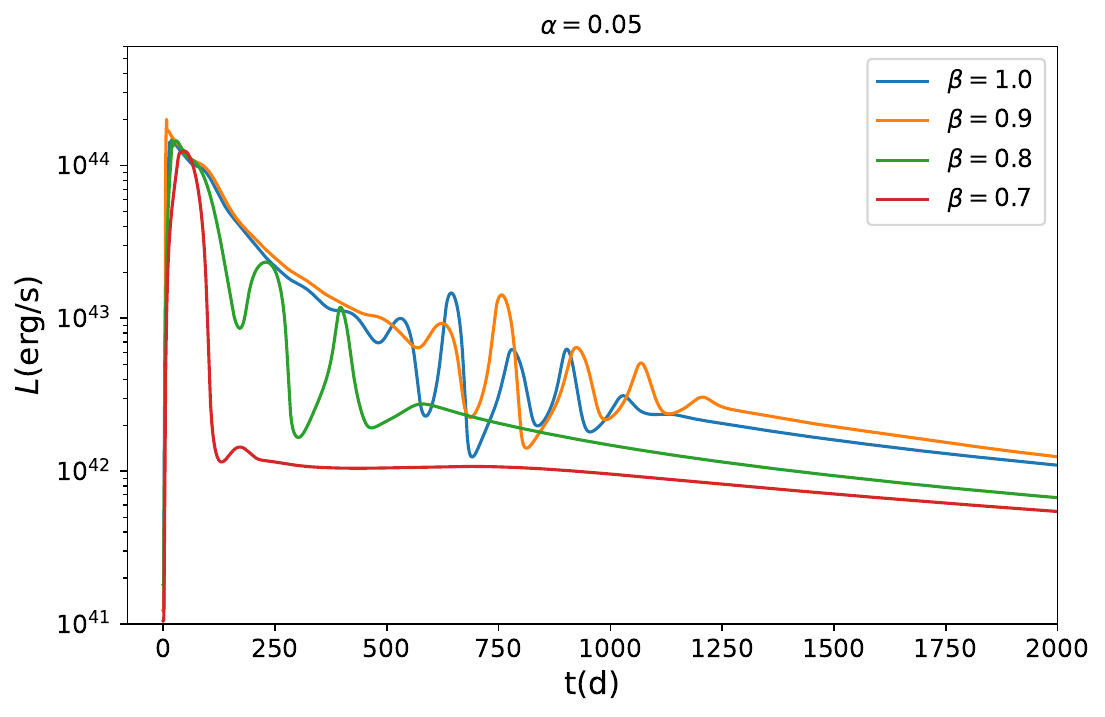}
\caption{ Light curves for different $\beta$, i.e., $\beta=1.0$, 0.9, 0.8, and 0.7. We set $M_{\rm{BH}}=10^{6}M_{\odot}$, $M_{\star}=0.1M_{\odot}$, $\mu=0.5$, $\alpha=0.05$, $f_{\rm{w}}=0$, and $R_{\rm{out}}=R_{\rm{c}}$ as equation (\ref{Rc}). 
}
\label{Lc_beta_0.1Msun_large_alpha}
\end{figure*}

\begin{figure*}[ht!]
\centering
\includegraphics[scale=0.6]{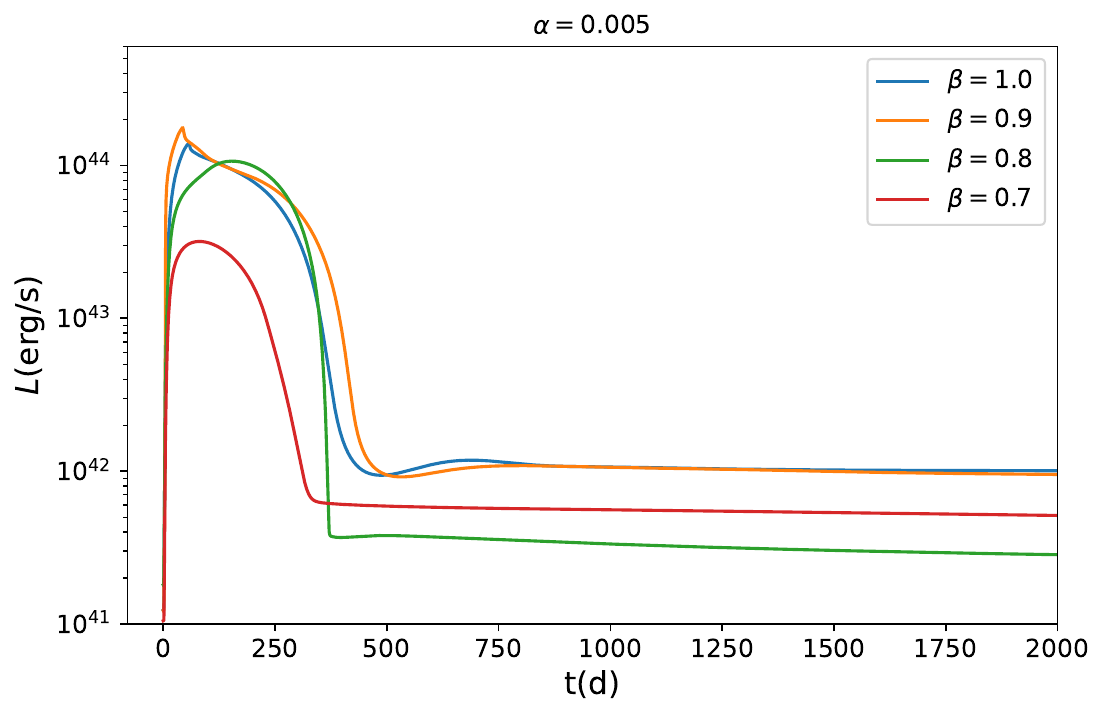}
\caption{ Same as Figure \ref{Lc_beta_0.1Msun_large_alpha} but with $\alpha=0.005$.
}
\label{Lc_beta_0.1Msun_small_alpha}
\end{figure*}

Taking the $\gamma=5/3$ fallback rate, in Figure \ref{Lc_beta_0.1Msun_large_alpha}, we plot the light curves calculated with $\alpha=0.05$ and different $\beta$, i.e., $\beta=1.0$, $0.9$, $0.8$, and $0.7$. We set $M_{\rm{BH}}=10^{6}M_{\odot}$, $M_{\star}=0.1M_{\odot}$, $\mu=0.5$, $f_{\rm{w}}=0$ and $R_{\rm{out}}=R_{\rm{c}}$ as equation (\ref{Rc}). In Figure \ref{Lc_beta_0.1Msun_small_alpha}, we show the light curves calculated with $\alpha=0.005$ and different $\beta$, i.e., $\beta=1.0$, $0.9$, $0.8$, and $0.7$. Other parameter settings are the same as in Figure \ref{Lc_beta_0.1Msun_large_alpha}. It can be seen that the effect of $\beta$ is quite similar to the $M_{\star}=M_{\odot}$ case shown in section \ref{Subsubsec 1Msun}. For a smaller star, the disk evolves faster, leading to an earlier onset and an earlier end of the radiation pressure instability. This is because a smaller star corresponds to a lower fallback rate (see equation (\ref{M_dot_fb_prop})) and a smaller size of the disk (see equation (\ref{Rc})). Consequently, the total luminosity is generally lower for a smaller star.

\vspace{1cm}
From the above calculation, it can be seen that the light curves show various behaviors in different parameter spaces. Based on the shape of the light curves, we find that the light curves in our calculation can be divided into three classes:

(i) light curves with oscillations caused by the radiation pressure instability;

(ii) light curves that drop significantly when the instability occurs and become flat in the late-time evolution;

(iii) light curves with the radiation pressure instability completely removed. 

Generally, type (i) and type (ii) light curves exist for $\mu >0.4$, where the radiation pressure instability plays a role in the disk and greatly affects the behaviors of the light curves. Type (iii) light curves occur for $\mu <0.4$, where the radiation pressure instability is completely removed from the disk. For $\mu >0.4$, type (ii) light curves appear for small $\alpha$ and small $\beta$, as well as large $R_{\rm{out}}$ (see section 3.3 in GQ26), all of which prevent the light curves of the disk from oscillating after the drop caused by the radiation pressure instability. In the following sections, we will show the application of type (ii) light curves to the optical/UV observations in TDEs and discuss the potential application of type (i) and type (iii) light curves in section \ref{Sec.Discussion}.

\section{Application to observations}
\label{Sec.Application to observations}

The optical/UV light curves of TDEs always show a rise-to-peak behavior followed by a quick decay in the early times, and many of them become flat in the late times. 
This long-term behavior is similar to the type (ii) light curves \footnote{Our classification of the calculated light curves is based on the behavior of the light curves after the initial peak. When applying our model to observation, we also include the rising phase of the light curves before the peak.} in our calculation.
So in this section, we study the applications of the type (ii) light curves to the optical/UV observations following a procedure similar to that in GQ26.

To account for the optical/UV emission in TDEs, simply a disk model is insufficient, as the optical/UV emission in TDEs could have different origins at different times. In the early times, since the estimated size of the optical/UV emission region is roughly 1–2 orders of magnitude larger than the theoretical circularization radius $R_{\rm{c}}$, these emissions are not expected to be directly powered by an accretion disk. Instead, these emissions could be powered by the shock induced by the collisions of debris streams before the formation of a circular accretion disk (\cite{Piran_2015}, \cite{Jiang2016ApJ}, \cite{Bonnerot2017MNRAS}), or they could be related to the reprocessing process in an optically thick envelope or disk outflows surrounding the disk (\cite{Ulmer1998A&A}, \cite{Ulmer1999ApJ}, \cite{Strubbe_Quataert_2009MNRAS}, \cite{Coughlin_2014}, \cite{Metzger_Stone2016MNRAS}, \cite{Roth_2016ApJ}, \cite{Dai2018ApJ}, \cite{Qiao2025MNRAS.539.3473Q}), both of which can absorb the high energy photons from the disk and re-emit in the optical/UV band. Recent simulations have shown that the pre-peak optical/UV emission is mainly powered by shocks, while the accretion can become competitive near or after the optical peak (\cite{Steinberg_2024Natur}, \cite{Huang2025}). Though our model cannot account for the collision process that may contribute to the optical/UV emission in the early times, by combining our disk model with a photosphere model and fitting the optical/UV light curves, we can give some constraints on when the accretion and the related reprocessing process starts to play a role in producing the optical/UV emission.   

In this work, we adopt a similar photosphere model to convert the disk luminosity into optical/UV flux as that in GQ26. In modeling the photosphere radius $R_{\rm{ph}}$, we first assume that $R_{\rm{ph}}$ can be described by a single power law function of the disk luminosity following \cite{Mockler2019ApJ} throughout the evolution, which can be written as
\begin{equation}
    R_{\rm{ph}} = R_{\rm{ph0}}(L/L_{\rm{Edd}})^{l}, \label{Rph}
\end{equation}
where $R_{\rm{ph0}}$ and $l$ are two parameters to be determined by fitting the light curves, and $L$ is the total luminosity of the disk. When this simple power law function of $R_{\rm{ph}}$ does not perform well during the fitting, we switch to the piecewise formula in modeling $R_{\rm{ph}}$ as equation (23) in GQ26. To maintain the continuity, we use the same photosphere model in fitting the long-term evolution of the optical/UV light curves. This usage of the photosphere model will be discussed in Section \ref{Subsec.The usage of the photosphere model}

Compared with GQ26, based on the adoption of a more realistic fallback rate and the upgrade of our disk model, our applications to the optical/UV light curves can have the following improvements: 

(1) We can fit the optical/UV light curves with reduced uncertainties around the optical peak. 
In GQ26, a parameter $t_{\rm{o}}$ has been introduced in fitting the light curves of ASASSN-15oi and ASASSN-14ae. These two sources are detected post-peak and $t_{\rm{o}}$ accounts for the uncertainties between the detection time and the time for the real optical peak. By incorporating the early rising phase of the fallback rate in our calculations, we can apply our model to TDEs with well-sampled optical light curves covering the evolution from the rising phase through the peak and the subsequent decay to the late-time flattening. For such TDEs, the parameter $t_{\rm{o}}$ can be removed from the fitting. 

(2) We can account for a wider range of optical/UV light curves in TDEs. As shown in section \ref{Subsec. effect of mu}, with the modified viscosity, the late-time luminosity can be enhanced to $\ge 10^{42}\rm{erg/s}$, which is much higher than the prediction of the traditional $\alpha$-viscosity. Thus, we can apply our model to sources with brighter late-time luminosity in optical/UV bands.

\subsection{Fitting procedure}
\label{Subsec. Fitting procedure}
In applying our model to the optical/UV data, the parameters concerning the time-dependent accretion disk, i.e., $M_{\rm{BH}}$, $M_{\star}$, $\beta$, $\mu$, $\alpha$, $f_{\rm{w}}$, cannot be constrained by fitting the optical/UV light curves, since this requires exploration of the full parameter space of our time-dependent disk model, which is not currently available. Instead, we can set some constraints on the bound of the model parameters based on observations, simulations and the effect of these parameters on the light curves as discussed in section \ref{Sec.Results}. Firstly, we take $M_{\rm{BH}}$ within the range given by the $M_{\rm{BH}}-\sigma_{\star}$ relation or the $M_{\rm{BH}}-M_{\rm{gal}}$ relation (\cite{Yao2023ApJL}, \cite{Leloudas_2019ApJ_AT2018dyb}). For $M_{\star}$, we mainly take small $M_{\star}$ to achieve small peak fallback rate. This is because the observed optical/UV light curves in TDEs often show a rapid decay following the initial rise to the peak. To account for this behavior with our model, a small peak fallback rate is required so that the steep decay caused by the radiation pressure instability can occur directly after the peak. Small $M_{\star}$ is also preferred by the fitting results of MOSFiT for many TDEs (\cite{Mockler2019ApJ}). For $\beta$, we mostly take $\beta_{\rm{c}}$, the critical value for the full disruption unless the observed light curve decays too steep to be fitted by light curves calculated with full disruption fallback rate. 
The value of $\mu$ is generally determined by the luminosity of the late-time plateau in the observed optical/UV light curves, as the late-time luminosity is primarily affected by $\mu$ for a given $\beta$. 
With a given $\mu$, $\alpha$ is tuned to have a suitable value to keep the light curves from oscillating. 
For $f_{\rm{w}}$, we take it in the range of $<0.5$, since the ratio of the mass loss rate in the outflows to $\dot{M}$ is found to be within this range in the simulation of a super-Eddington accretion disk around a super-massive BH(\cite{Jiang2014ApJ.796.106J}). Within these physical limits of the parameters, we calculate several disk light curves with different sets of $M_{\rm{BH}}$, $M_{\star}$, $\beta$, $\mu$, $\alpha$, and $f_{\rm{w}}$. Then we fit them to the observed optical/UV light curves with the photosphere model, and determine the two photosphere parameters $R_{\rm{ph0}}$ and $l$ via fitting. By comparing the fitting results, we can determine a best set of parameters for the time-dependent disk. This procedure will be discussed in section \ref{Subsec. Exploration of the full parameter space}.

We use the optical and UV light curve data collected from the manyTDE repository \footnote{https://github.com/sjoertvv/manyTDE} (\cite{Mummery2024MNRAS}). The data are host-subtracted. We correct the Galactic extinction following the sample code presented with the manyTDE repository. In this work, we mainly consider TDEs with a clear detection of the optical peak. Since we find it difficult to fit the light curves in both the optical and UV bands simultaneously with our model, we fit only the band with more observational data.

\subsection{Fitting results}
\label{Subsec. Fitting results}
\begin{figure*}[ht!]
\centering
\includegraphics[scale=0.4]{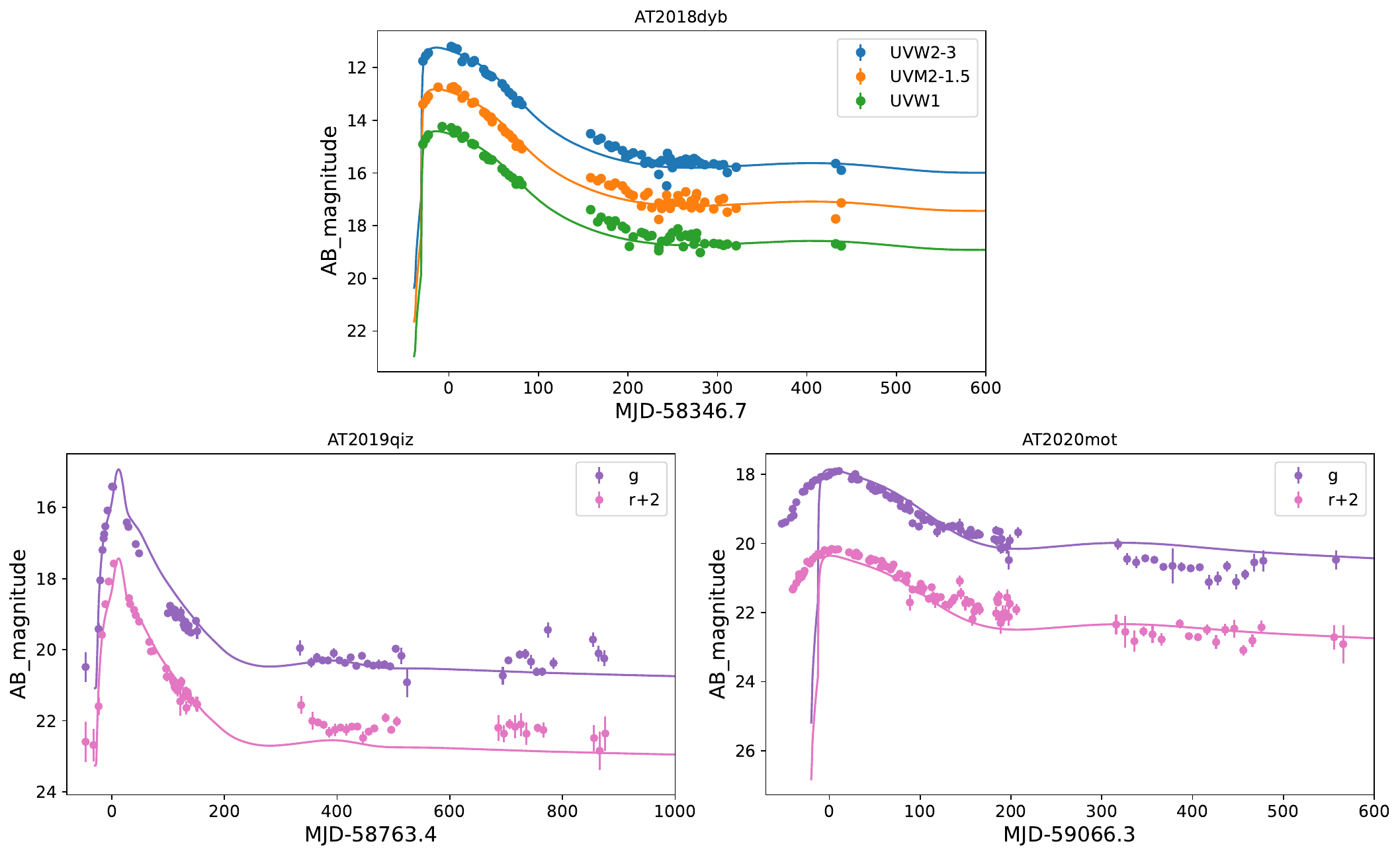}
\caption{The best fit optical/UV light curves of AT2018dyb, AT2019qiz, and AT2020mot. The solid light curves are first calculated by our disk model with parameters listed in Table \ref{disk parameters in fitting}, and fitted to the observed data with the photosphere model. All the light curves are plotted with the optical peak set at $t=0$.  }
\label{fitting_2}
\end{figure*}

In Figure \ref{fitting_2}, we show the optical/UV light curves of three TDEs which can be fitted by our type (ii) light curves, including AT2018dyb, AT2019qiz and AT2020mot. The data are plotted with the optical peak shifted to $t=0$ on the x-axis. The light curves calculated with our model are plotted by aligning the peak in the light curves with the optical peak in the observations. The parameters in calculating the light curves of the time-dependent disk and the best-fit parameters for the photosphere are listed in Table \ref{disk parameters in fitting}. 
It can be seen in Figure \ref{fitting_2} that the disk light curves in our calculations rise to peak very quickly, thus our model cannot well fit the slow rise in the optical light curves of AT2020mot. Thus for this source, we use only the post-peak data during the fitting. Our model provides a good fit for all three TDEs from a time near the optical peak. This indicates that the early rising stage in the optical bands may not be intrinsically powered by the accretion disk or its related reprocessing process, while accretion begins to contribute to the optical/UV emission from the optical peak, consistent with current findings in simulations (\cite{Steinberg_2024Natur}, \cite{Huang2025}).

From the optical peak, the long-term behavior of the optical/UV light curves can be well described by the continuous evolution of the accretion disk. Here, as applications of our type (ii) light curves, the decay-to-flatten behavior of these three TDEs is interpreted as a result of radiation pressure instability in the disk. Specifically, the decay level of the optical/UV flux reflects the severity of the radiation pressure instability. From the optical peak to the late-time plateau, the change in single-band magnitude is 4.8 (in g band), 4.4 (in UVW1 band), and 3.1 (in g band) for AT2019qiz, AT2018dyb and AT2020mot, respectively. Correspondingly, the value of $\mu$ in calculating the disk evolution is chosen to be 1.0, 0.6, and 0.4 for these three TDEs, as shown in Table \ref{disk parameters in fitting}. For smaller $\mu$, the radiation pressure instability is more strongly suppressed, thus the decay magnitude in optical/UV band is smaller. The $\mu=0.4$ case we used for AT2020mot is a critical case of type (ii) light curve. For TDEs with smaller decay in optical/UV band than AT2020mot, we have to take $\mu<0.4$ to interpret their optical/UV light curves, which transition to our type (iii) light curves. The potential application of type (iii) light curves will be discussed in section \ref{Subsec. Other potential applications}.   

\begin{figure*}[ht!]
\centering
\includegraphics[scale=0.4]{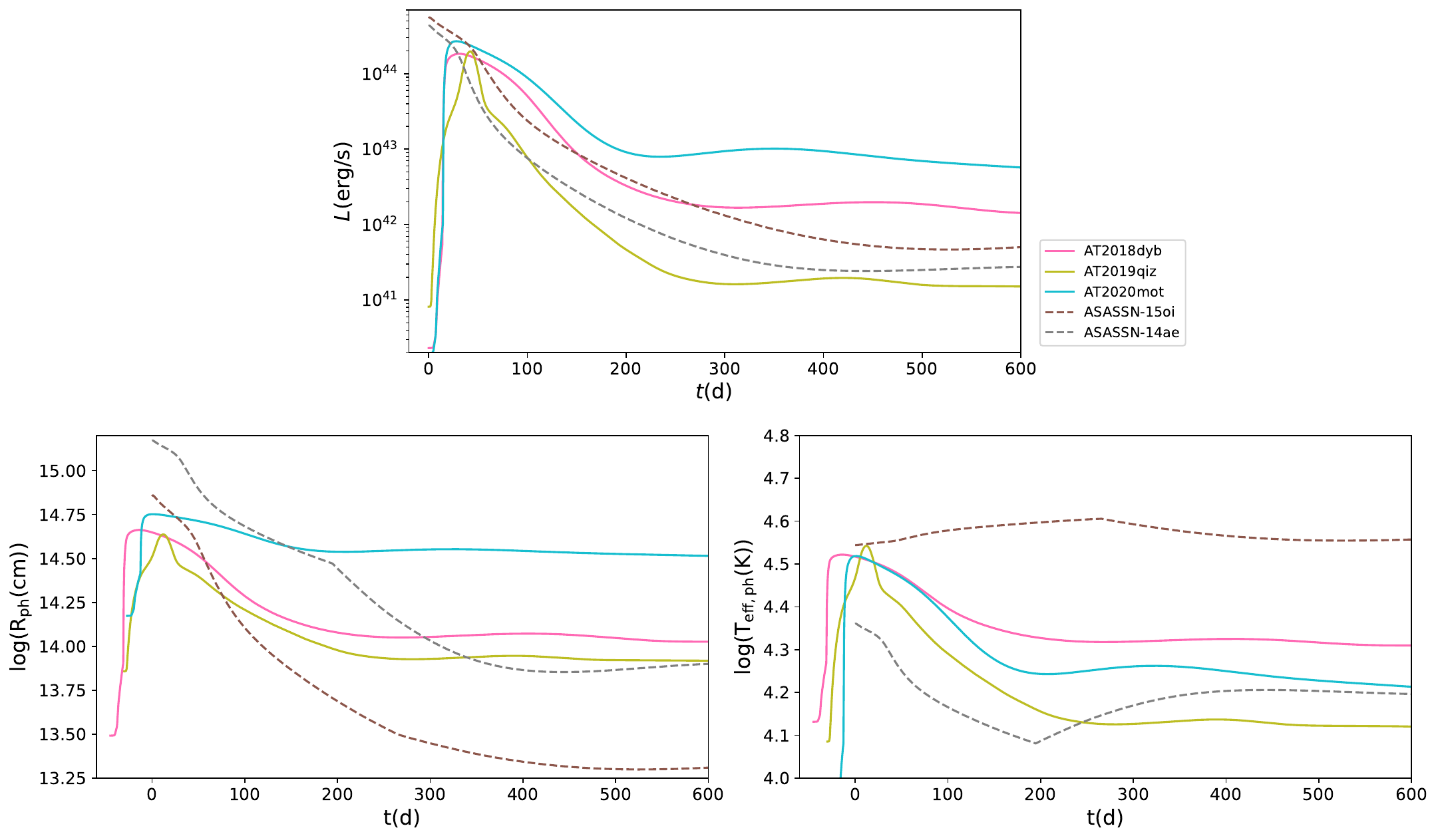}
\caption{Upper panel: the light curves of the disk used to fit the optical/UV observations in Figure \ref{fitting_2} (solid lines) and in GQ26 (dashed lines). Lower panels: the photosphere radius $R_{\rm{ph}}$ and effective temperature of the photosphere $T_{\rm{eff, ph}}$ obtained from fitting.  }
\label{Rph_Teff}
\end{figure*}

In Figure \ref{Rph_Teff}, we show the disk light curves, the evolution of $R_{\rm{ph}}$ and the effective temperature of the photosphere $T_{\rm{eff,ph}}$, respectively. Besides the three TDEs we fit in this paper (solid lines), we also plot the results of ASASSN-15oi and ASASSN-14ae that we fit in GQ26 (dashed lines) to better demonstrate the application of our model. The disk light curves in the upper panel show that, the decay magnitudes in this work vary over a wider range with different choice of $\mu$, compared to those in GQ26 with fixed $\mu=1.0$. In the lower two panels, it can be seen that the value of $R_{\rm{ph}}$ and $T_{\rm{eff,ph}}$ we get from our fitting of the photosphere model are comparable to those that have been derived from fitting blackbodies to the photometry (\cite{Hammerstein2022AAS}, \cite{Leloudas_2019ApJ_AT2018dyb}, \cite{Holoien2016MN_ASASSN-15oi_1}), with peak $R_{\rm{ph}}$ between $10^{14.2}-10^{15.2}\rm{cm}$, and peak $T_{\rm{eff,ph}}$ between $10^{4.2}-10^{4.8}\rm{K}$. Regarding the evolutionary trend, all TDEs show a decrease in $R_{\rm{ph}}$ after peak, while the evolution of $T_{\rm{eff,ph}}$ varies across different events. 
The change in $T_{\rm{eff,ph}}$ after the initial peak is generally within $\sim 0.3\rm{dex}$, except for AT2019qiz that shows a relatively severe decay with $\sim 0.4\rm{dex}$.
This indicates that most of our results are roughly consistent with the photosphere properties inferred from observation, which show a mild evolution of $T_{\rm{eff,ph}}$. However, in order to examine whether our model can better reproduce these observational properties, a more extensive exploration of the parameter space is required, which we will further investigate in future work.   

\begin{deluxetable*}{l c c c c c c | c c}
\tablecaption{Parameters for the time-dependent disk (column 2 - 7) and best-fit parameters for the photosphere (column 8 and 9) \label{disk parameters in fitting}}
\tablehead{
\colhead{TDE} & \colhead{$\beta$} & \colhead{$M_{\rm{BH}}(M_{\odot})$} & \colhead{$M_{\star}(M_{\odot})$} & \colhead{$\mu$} & \colhead{$\alpha$} & \colhead{$f_{\rm{w}}$} & \colhead{$R_{\rm{ph0}}(\rm{cm})$} & \colhead{$l$} 
}
\startdata
AT2019qiz & 0.75 & $2\times 10^{6}$  & 0.1 & 1.0 & 0.0035 & 0.0 & $10^{14.66}$ & 0.23 \\
AT2018dyb & 0.9  & $7\times 10^{6}$  & 0.2 & 0.6 & 0.011  & 0.4 & $10^{14.87}$ & 0.30 \\
AT2020mot & 0.9  & $7.5\times 10^{6}$  & 0.2 & 0.4 & 0.05 & 0.2 & $10^{14.83}$ & 0.14 \\
\enddata
\end{deluxetable*}

\section{Discussion}
\label{Sec.Discussion}

\subsection{The evolution of the photosphere}
\label{Subsec.The usage of the photosphere model}
In fitting the optical/UV light curves in TDEs, we use the photosphere model described in Section \ref{Subsec. Fitting procedure} to convert the disk luminosity into multi-band flux. This is motivated by the early-time observations in TDEs (e.g., \cite{Velzen2020SSRv}, \cite{Gezari2021_annurev}), and is also consistent with simulation of the super-Eddington accretion in the early phase of TDEs (\cite{Dai2018ApJ}, \cite{Qiao2025MNRAS.539.3473Q}). 
We apply the disk plus photosphere model to describe the long-term evolution of the optical/UV light curves observed in TDEs as a whole. As shown in Figure \ref{Rph_Teff}, we find that in the early times, $R_{\rm{ph}}$ is about two orders of magnitude larger than $R_{\rm{c}}$. Then $R_{\rm{ph}}$ recedes with decreasing luminosity, but is still larger than $R_{\rm{c}}$ in the late times.
On the other hand, a spreading disk model has been proposed to study the late-time optical/UV plateau (\cite{Mummery2024MNRAS}, \cite{Mummery2025MNRAS.541..429M_Ref6}, \cite{Alush_2025ApJ_1}, \cite{Alush_2025_2}, \cite{Guolo2025arXiv_Compact_Accretion_Disks}).  
In our work, if we only consider the disk emission, the optical/UV flux will be lower than that in observations. This is because we have fixed the outer radius of the disk at $R_{\rm{c}}$ in our calculation, which makes the emission region too small to account for the observed emission. 
Nevertheless, the photosphere we adopt in our fitting provides a phenomenologically equivalent description of the observed features as the spreading disk model.
Our disk plus photosphere model provides a unified framework for describing the long-term evolution of the optical/UV light curves, i.e., from the early-time rise through the post-peak decay, to the late-time flattening observed in TDEs.

In our model, when the light curve decays steeply, the disk is experiencing the radiation pressure instability, and quickly transitions from a slim disk to a thin disk. During this time, the total luminosity of the disk decreases, and the super-Eddington wind ceases, causing the photosphere to recede. In the late times, $T_{\rm{eff}}$ of the disk changes very little, producing the flat feature in the type (ii) light curves. In this way, the long-term behavior of the optical/UV light curves can be interpreted with the evolution of a time-dependent disk plus a photosphere. 
However, this application is based on the assumption that the evolution of the photosphere follows that of the disk (see equation (\ref{Rph})), so that the optical/UV emission from the photosphere can intrinsically reflect the luminosity change in the disk. In TDEs, the detailed evolution of the photosphere could be complicated, and may not be dominated by the evolution of the disk. 
For example, the early-time evolution of the optical/UV emission could be powered by the strong shock in collisions, and do not reflect the evolution of the accretion disk. 

\subsection{Other potential applications}
\label{Subsec. Other potential applications}
In this work, we have only demonstrated some applications of the type (ii) light curves in the optical/UV band. The other two types of light curves also have potential application in TDE observations. 
Firstly, the type (iii) light curves with the instability completely removed are largely consistent with those investigated in previous works (e.g. \cite{Mummery_Balbus2020MNRAS}, \cite{Alush_2025ApJ_1}). And this type of light curves could be applied to explain the smooth decay of some X-ray light curves in TDEs such as ASASSN-14li (\cite{Bright_2018_ASASSN-14li}, \cite{Mummery_Balbus2020MNRAS}). Besides, some optical/UV light curves show a continuous decay but not a true plateau in the late times (\cite{Alush_2025_2}), which is reminiscent of the behaviors of our type (iii) light curves. 

Some TDEs show large-magnitude variability such as multiple flares in their light curves (e.g. \cite{Auchettl2017ApJ}, \cite{Guolo2024ApJ_op_X_unification}, \cite{Payne2021ApJ_ASASSN14ko}, (\cite{Liu_2023AA_eRASSt_J0456}). These observed features generally have different explanations in previous works, including super-massive BH binaries (\cite{Liu_2014ApJ_SDSSJ1201}, \cite{Shu_2020NatCo}), repeating partial TDEs (\cite{Liu_2023AA_eRASSt_J0456}, \cite{Liu_2024AA_eRASSt_J0456}, \cite{Payne2021ApJ_ASASSN14ko}, \cite{Huang2023ApJ_ASASSN14ko}), Lense–Thirring precession (\cite{AT2020ocn_Pasham2024Natur}), etc. From the results shown in section \ref{Sec.Results}, we find that our type (i) light curves can also reproduce similar features to some of these TDEs. Thus, the radiation pressure instability serves as a potential explanation for this large-magnitude variability. In future work, a more detailed comparison of our model predictions with the observed data is necessary to investigate this potential application.

\subsection{Exploration of the full parameter space}
\label{Subsec. Exploration of the full parameter space}
As stated in section \ref{Subsec. Fitting procedure}, when fitting the observed optical/UV light curves with our model, we separately treat the time-dependent accretion disk and the photosphere model. This is because the global fitting for data with these two models requires full explorations of the parameter space of the time-dependent disk model, which demands a great amount of calculations and is not currently available. In future work, it is necessary to fully explore the parameter spaces and build a library to store all our calculated light curves. As shown in section \ref{Sec.Results}, the behaviors of our calculated light curves vary significantly with different parameter settings, thus, the parameter spaces should be explored in as much detail as possible. In this way, the disk light curve for any arbitrary set of parameters can be obtained via interpolation from the library. This will allow us to fit the observed light curves globally and better constrain all the model parameters.


\section{Conclusion}
\label{Sec.Conclusion}
In this work, we studied the light curves of the time-dependent disk in the environment of TDEs, using an updated model based on the framework proposed by GQ26. In terms of the disk model, we incorporate the effect of the modified viscosity and disk wind, which are represented by two new parameters $\mu$ and $f_{\rm{w}}$, respectively. Besides, we adopt a more realistic fallback rate that includes the initial rising phase for the mass injection to the disk. 
We systematically test the effects of all the newly incorporated parameters in our time-dependent disk model on the light curves, including $\mu$, $f_{\rm{w}}$ and the impact parameter $\beta$. We find that:

\begin{itemize}
    \item $\mu$ can shape the unstable region of the S-shaped curve and influence the occurrence of radiation pressure instability. For $\mu<0.4$, the instability is completely removed from the disk, and the light curves always evolve stably. For $\mu>0.4$, the light curves are affected by the instability, and the variation amplitudes are smaller for smaller $\mu$. In addition, we find that the oscillation amplitudes can decay over time in some of our calculations with $\mu=0.5$. 
    \item The disk wind mainly affects the time for the radiation pressure instability to occur. With stronger wind, i.e., larger $f_{\rm{w}}$, the radiation pressure instability occurs earlier.
    \item $\beta$ does not influence the existence of the radiation pressure instability in the disk, but it affects the presence of oscillations in the light curves. For $\beta$ smaller than the critical value of the full disruption, the oscillations are increasingly suppressed as $\beta$ decreases. Below a certain value, the oscillations are completely suppressed. In this case, the light curve exhibits a decay-to-flatten behavior similar to that previously found for small $\alpha$, but the decay could be quite steep for small $\beta$ due to the rapid decline of the fallback rate. 
\end{itemize}

Based on the shape of the light curves obtained from our calculation, we find that the light curves can be generally divided into three types: (i) light curves with oscillations caused by the radiation pressure instability, which occurs for $\mu >0.4$ and large $\alpha$; (ii) light curves that drop steeply when the radiation pressure instability happens and become flat in the late times, which occurs for $\mu >0.4$ and small $\alpha$, small $\beta$ as well as large $R_{\rm{out}}$ (see Figure 5 in GQ26); (iii) light curves with the radiation pressure instability completely removed, which occurs for $\mu \le 0.4$.

We find that our type (ii) light curves are reminiscent of some optical/UV light curves observed in TDEs. Together with a photosphere model, we fit the optical/UV light curves of some TDEs with our type (ii) light curves. We find that the long-term decay-to-flatten behaviors of the optical/UV light curves for some TDEs can be well described by our model. In particular, we can model TDEs with different magnitudes of decline, which can be significantly modulated by the value of $\mu$ in our disk model. 


\begin{acknowledgments}
We thank the organizers and participants of the TDE Full Process Simulation Seminar Series for the valuable lectures and discussions. This work is supported by the Strategic Priority Research Program of the Chinese 
Academy of Science (Grant No. XDB0550200),
National Key R\&D Program of China (No. 2023YFA1607903) and National Natural Science Foundation of China (Grant No. 12333004, 12173048).
\end{acknowledgments}

\appendix
\section{Time-dependent disk equations with outflows}
\label{appsec:Time-dependent disk equations with outflows}
We consider an axisymmetric time-dependent accretion disk in cylindrical coordinates. The continuity equation is 
\begin{equation}
    \frac{\partial \rho}{\partial t} + \frac{\partial }{r\partial r}(r\rho v_{r}) + \frac{\partial (\rho v_{\rm{z}})}{\partial z} = 0 .  \label{app:continuity_rho}
\end{equation}
By integrating equation (\ref{app:continuity_rho}) over z-direction, we obtain
\begin{equation}
    \frac{\partial \Sigma}{\partial t} + \frac{\partial }{r\partial r}(r\Sigma  v_{r}) + 2\rho_{\rm{e}}v_{\rm{w}} = 0,  \label{app:continuity_Sigma}
\end{equation}
where the term $2\rho_{\rm{e}}v_{\rm{w}}$ represents the mass loss rate in outflows per unit disk surface area. This term is an approximation of the integration $\int_{-H}^{H} \frac{\partial (\rho v_{z})}{\partial z}dz$. Here we have used $\rho_{\rm{e}}$ and defined an average wind velocity $v_{\rm{w}}$ to approximate this vertical mass loss rate. 

The equation of motion in $\varphi$-direction is
\begin{equation}
    \frac{\partial v_{\varphi}}{\partial t} + v_{r}\frac{\partial v_{\varphi}}{\partial r} + v_{z}\frac{\partial v_{\varphi}}{\partial z} + \frac{v_{r}v_{\varphi}}{r} = \frac{1}{\rho r^{2}}\frac{\partial}{\partial r}(r^{2}t_{r\varphi}), \label{app:vphi}
\end{equation}
where $v_{\varphi}=r\Omega$ is the velocity in $\varphi$-direction. Combining equation (\ref{app:vphi}) with equation (\ref{app:continuity_rho}), we obtain the angular momentum equation
\begin{equation}
    \frac{\partial}{\partial t}(\rho rv_{\varphi}) + \frac{1}{r}\frac{\partial}{\partial r}(r^{2}\rho v_{r}v_{\varphi}) + \frac{\partial}{\partial z}(r\rho v_{\varphi}v_{z}) = \frac{\partial}{r\partial r}(r^{2}t_{r\varphi}), \label{app:angular momentum_rho}
\end{equation}
where $t_{r\varphi}$ is the $r\varphi$ component of the stress tensor. 
By integrating equation (\ref{app:angular momentum_rho}) over z-direction, we obtain
\begin{equation}
    \frac{\partial}{\partial t}(\Sigma rv_{\varphi}) + \frac{1}{r}\frac{\partial}{\partial r}(r^{2}\Sigma v_{r}v_{\varphi}) + 2\rho_{\rm{e}}v_{\rm{w}}rv_{\varphi} = \frac{\partial}{r\partial r}(r^{2}T_{r\varphi}). \label{app:angular momentum_Sigma}
\end{equation}
Here we have used similar approximation in integrating the $\frac{\partial}{\partial z}$ term as in deriving equation (\ref{app:continuity_Sigma}).

The conservation equation of thermal energy can be written as
\begin{equation}
    \frac{\partial}{\partial t}(\rho e) + \frac{1}{r}\frac{\partial}{\partial r}(r\rho ev_{r}) + \frac{\partial}{\partial z}(\rho ev_{z}) + \frac{p}{r}\frac{\partial (rv_{r})}{\partial r} = q^{+}_{\rm{vis}} - \frac{\partial F_{\rm{rad}}}{\partial z}, \label{app:energy_rho}
\end{equation}
where $q^{+}_{\rm{vis}}$ is the viscous heating per unit volume, $F_{\rm{rad}}$ is the radiative flux per unit surface area. We have assumed the term $\frac{\partial v_{z}}{\partial z}$ to be negligible in deriving equation (\ref{app:energy_rho}) following \cite{Feng_2019ApJ}.
By integrating equation (\ref{app:energy_rho}) over z-direction, we obtain
\begin{equation}
    \frac{\partial E}{\partial t} + \frac{\partial(rE v_{\rm{r}})}{r\partial r} + 2\rho_{\rm{e}}e_{\rm{e}}v_{\rm{w}} + \Pi\frac{\partial(r v_{\rm{r}})}{r\partial r}= Q^{+}_{\rm{vis}} - Q^{-}_{\rm{rad}}, \label{app:energy_E}
\end{equation}
where we have used similar approximation in integrating the $\frac{\partial}{\partial z}$ term as in deriving equation (\ref{app:continuity_Sigma}).

By defining $\dot{\Sigma}_{\rm{out}} = 2\rho_{\rm{e}}v_{\rm{w}}$ and $Q^{-}_{\rm{out}} = 2\rho_{\rm{e}}e_{\rm{e}}v_{\rm{w}}$, and substituting them into equation (\ref{app:continuity_Sigma}), (\ref{app:angular momentum_Sigma}) and (\ref{app:energy_E}), we can obtain equation (\ref{mass conserv}), (\ref{angular momentum conserv}) and (\ref{energy conserv2}).

\section{Steady disk equations with outflows}
\label{appsec:Steady disk equations with outflows}
In our time-dependent disk model, we also need the steady disk solutions for setting the initial conditions and boundary conditions. To calculate the steady disk solution, we first need to specify $\dot{M}$ with a given mass injection rate $\dot{M}_{\rm{inject}}$. For a given $\dot{M}_{\rm{inject}}$ at $R_{\rm{out}}$, the following mass conservation should hold (\cite{Takeuchi_Mineshige2009}, \cite{Feng_2019ApJ})
\begin{equation}
    \dot{M}(r) + \dot{M}_{\rm{out}}(r) = \dot{M}_{\rm{inject}}.  \label{app:M_dot_inject_conserv}
\end{equation}
Similar to Section \ref{Subsec.Disk model}, we set $\dot{M}_{\rm{out}}(r)=f_{\rm{w}} \beta_{\rm{rad}}(r)\dot{M}_{\rm{inject}}$. Thus, $\dot{M}(r)$ can be written as 
\begin{equation}
    \dot{M}(r)=\dot{M}_{\rm{inject}} - \dot{M}_{\rm{out}}(r)=(1-f_{\rm{w}}\beta_{\rm{rad}}(r))\dot{M}_{\rm{inject}}. \label{app:M_dot}
\end{equation}

We can then obtain the stable solution of the accretion disk by solving the local energy balance equation.
For a steady disk, taking $\partial S/\partial t = 0$, the energy loss rate through horizontal advection can be expressed as 
\begin{equation}
    Q^{-}_{\rm{adv}} = \Sigma Tv_{\rm{r}}\frac{d S}{dr}=\frac{\dot{M}}{2\pi r^{2}}\frac{p}{\rho}\xi, \label{app:Q_adv_steady}
\end{equation}
where $\xi$ is assumed to be a constant. We take $\xi=1.5$ throughout this work following \cite{Watarai2006ApJ}. 
The energy conservation equation for a steady disk can then be written as
\begin{equation}
    2D(r)= 2C_{2}\frac{4\sigma T^{4}}{3\kappa \rho H}  + \frac{\dot{M}}{2\pi r^{2}}\frac{p}{\rho}\xi. \label{app:steady energy balance}
\end{equation}
where $D(r)$ is the viscous dissipation rate per unit disk face area in a steady disk, and can be expressed as, 
\begin{equation}
  D(r)=\frac{3GM_{\rm{BH}}\dot{M}(r)}{8\pi r^{3}}g(r). \label{app:D(r)}
\end{equation}
Here 
$g(r)=1-\sqrt{R_{\rm{ISCO}}/r}$. 

In equation (\ref{app:steady energy balance}), the energy loss through outflows has been incorporated in the viscous heating term and the advection term through $\dot{M}$, which is calculated with equation (\ref{app:M_dot}).
Combining equation (\ref{app:Q_adv_steady}), (\ref{app:steady energy balance}) and (\ref{app:D(r)}) together with (\ref{total pressure}), (\ref{hydrostatic equilibrium}) and (\ref{Q_vis}), we obtain the solutions for the local steady disk with Newton's iteration method. When calculating the initial condition of the time-dependent disk, first we set $\dot{M}(r)=\dot{M}_{\rm{inject}}$ to obtain a set of steady disk solution, and calculate $\beta_{\rm{rad}}(r)$. Then we calculate $\dot{M}(r)$ with equation (\ref{app:M_dot}), and obtain a new set of steady disk solution, which we adopt as the initial condition.

\begin{figure*}[ht!]
\centering
\includegraphics[scale=0.4]{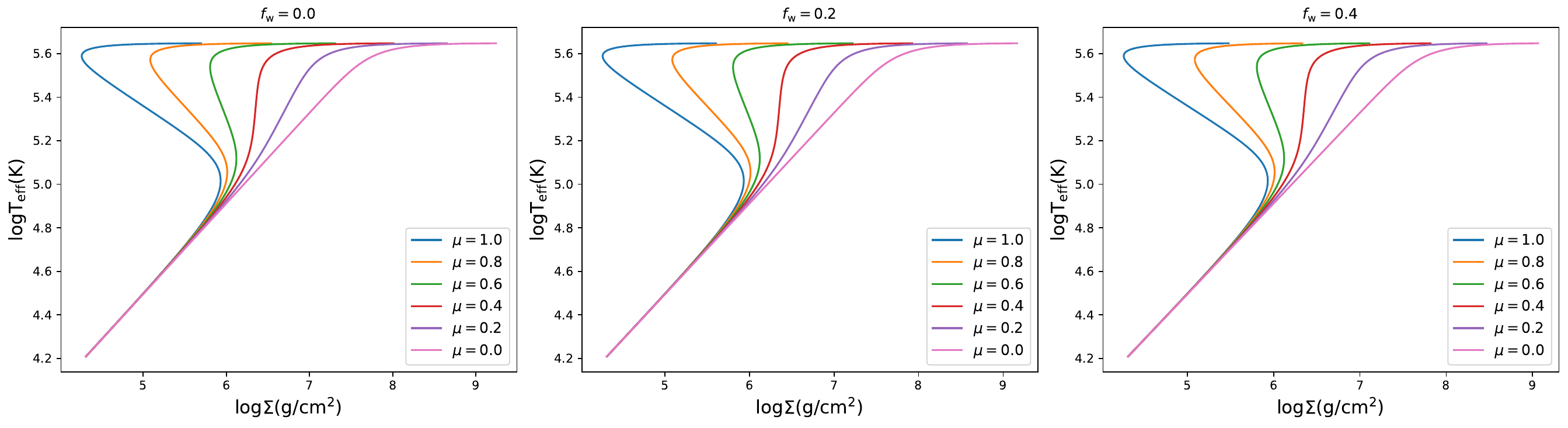}
\caption{Local stationary solution calculated with $M_{\rm BH}=10^{6}M_{\odot}$and $\alpha=0.01$ for different $\mu$ at $10R_{\rm{S}}$
with a fixed $f_{\rm{w}}$. 
The left, middle and right panels are for $f_{\rm{w}}=0.0$, $f_{\rm{w}}=0.2$ and $f_{\rm{w}}=0.4$, respectively.}
\label{Subplot_Scurve_mu}
\end{figure*}

In Figure \ref{Subplot_Scurve_mu}, we show the solutions on the $T_{\rm{eff}}-\Sigma$ plane at $10R_{\rm{S}}$ for different $\mu$ with a fixed $f_{\rm{w}}$. We take $M_{BH}=10^{6}M_{\odot}$ and $\alpha=0.01$. The left, middle and right panels are for $f_{\rm{w}}=0.0$, $f_{\rm{w}}=0.2$ and $f_{\rm{w}}=0.4$, respectively. It can be seen that $f_{\rm{w}}$ does not obviously affect the S-shaped curve (consistent with previous findings in \cite{Wu_2022ApJ_outflow}), while $\mu$ greatly affects the upper and middle branches of the S-shaped curves, which are dominated by radiation pressure in the disk. The negative slope on the S-shaped curve becomes smaller with smaller $\mu$, and disappears for $\mu<0.4$, implying that the radiation pressure instability is completely suppressed for $\mu<0.4$. This is equally true for different values of $f_{\rm{w}}$ as shown in Figure \ref{Subplot_Scurve_mu}.

\bibliography{main}{}
\bibliographystyle{aasjournalv7}



\end{document}